\documentclass{aa}  
\usepackage{amssymb} 

\usepackage{lscape} 
\usepackage{placeins} 
\usepackage{stfloats} 

\usepackage[switch, modulo]{lineno} 

\usepackage{natbib,twoopt}
\usepackage[breaklinks=true]{hyperref} 
\bibpunct{(}{)}{;}{a}{}{,} 
\makeatletter

\newcommandtwoopt{\citeads}[3][][]{\href{http://scixplorer.org/abs/#3}%
{\def\hyper@linkstart##1##2{}%
\let\hyper@linkend\@empty\citealp[#1][#2]{#3}}}
\newcommandtwoopt{\citepads}[3][][]{\href{http://scixplorer.org/abs/#3}%
{\def\hyper@linkstart##1##2{}%
\let\hyper@linkend\@empty\citep[#1][#2]{#3}}}
\newcommandtwoopt{\citetads}[3][][]{\href{http://scixplorer.org/abs/#3}%
{\def\hyper@linkstart##1##2{}%
\let\hyper@linkend\@empty\citet[#1][#2]{#3}}}
\newcommandtwoopt{\citeyearads}[3][][]%
{\href{http://scixplorer.org/abs/#3}
{\def\hyper@linkstart##1##2{}%
\let\hyper@linkend\@empty\citeyear[#1][#2]{#3}}}

\makeatother

\usepackage{graphicx}
\usepackage{txfonts}

\newcommand{\cmark}{\checkmark}
\newcommand{\xmark}{--} 

\begin{document} 

    \title{The complex inner disk of the \mbox{Herbig Ae} star HD~100453 with VLTI/MATISSE}

    \author{L.N.A. van~Haastere\inst{1}
\and
J. Varga\inst{2,3,1}
\and
M.R. Hogerheijde\inst{1,4}
\and
C. Dominik\inst{4}
\and
M. Scheuck\inst{5}
\and
A. Matter\inst{6}
\and
R. van~Boekel\inst{5}
\and
B. Lopez\inst{6}
\and
M. Abello\inst{6}
\and
J.-C. Augereau\inst{7}
\and
P. Boley\inst{5}
\and
W.-C. Danchi\inst{8}
\and
V. G\'amez~Rosas\inst{9}
\and
Th. Henning\inst{5}
\and
K.-H. Hofmann\inst{10}
\and
M. Houll\'e\inst{6}
\and
W. Jaffe\inst{1}
\and
J. Kobus\inst{11}
\and
E. Kokoulina\inst{9}
\and
L.H. Leftley\inst{6}
\and
M. Letessier\inst{7}
\and
J. Ma\inst{7}
\and
F. Millour\inst{6}
\and
E. Pantin\inst{12}
\and
P. Priolet\inst{7}
\and
D. Schertl\inst{10}
\and
J. Scigliuto\inst{6}
\and
G. Weigelt\inst{10}
\and
S. Wolf\inst{11}
\and
P. Berio\inst{6}
\and
F. Bettonvil\inst{14}
\and
P. Cruzal\`ebes\inst{6}
\and
M. Heininger\inst{10}
\and
J.W. Isbell\inst{15}
\and
S. Lagarde\inst{6}
\and
A. Meilland\inst{6}
\and
R. Petrov\inst{6}
\and
S. Robbe-Dubois\inst{6}
\and
MATISSE Collaboration
}

\institute{
Leiden Observatory, Leiden University, Einsteinweg 55, 2333 CC Leiden, The Netherlands
\and
Konkoly Observatory, Research Centre for Astronomy and Earth Sciences, HUN-REN, Konkoly-Thege Mikl\'os \'ut 15-17, 1121 Budapest, Hungary
\and
CSFK, MTA Centre of Excellence, Konkoly-Thege Mikl\'os \'ut 15-17, H-1121 Budapest, Hungary
\and
Anton Pannekoek Institute for Astronomy, University of Amsterdam, the Netherlands
\and
Max Planck Institute for Astronomy, K\"onigstuhl 17, D-69117 Heidelberg, Germany
\and
Universit\'e C\^ote d’Azur, Observatoire de la C\^ote d’Azur, CNRS, Laboratoire Lagrange, France
\and
Univ. Grenoble Alpes, CNRS, IPAG, 38000 Grenoble, France
\and
NASA Goddard Space Flight Center, Astrophysics Division, Greenbelt, MD 20771, USA
\and
STAR Institute, University of Li\`ege, Li\`ege, Belgium
\and
Max-Planck-Institut f\"ur Radioastronomie, Auf dem H\"ugel 69, 53121, Bonn, Germany
\and
Institute of Theoretical Physics and Astrophysics, University of Kiel, Leibnizstr. 15, 24118 Kiel, Germany
\and
AIM, CEA, CNRS, Universit\'e Paris-Saclay, Universit\'e Paris Diderot, Sorbonne Paris Cit\'e, F-91191 Gif-sur-Yvette, France
\and
Institute for Mathematics, Astrophysics and Particle Physics, Radboud University, P.O. Box 9010, MC 62 NL-6500 GL Nijmegen, the Netherlands
\and
SRON Netherlands Institute for Space Research, Niels Bohrweg 4, 2333 CA Leiden, The Netherlands
\and
Steward Observatory, University of Arizona, 933 N. Cherry Avenue, Tucson, AZ 85721, USA
}

\date{\today}
   
  \abstract 
   {
   The inner regions of planet-forming disks hold invaluable insights for our understanding of planet formation. The inner disk regions that might be affected by already formed planets are of particular interest. The disk around the Herbig star HD~100453 presents one such environment, with an inner disk that is significantly misaligned with respect to the outer disk. }
   {
   This paper expands the existing H band (PIONIER) and K band (GRAVITY) interferometric studies of the inner disk of HD~100453 to the L band with the MATISSE VLTI instrument. Based on snapshot data spanning approximately four years, we aim to understand the inner disk structures and their potential time evolution better.}
  {
   Based on the MATISSE data we obtained, we used a combination of analytical models and image reconstruction to constrain the disk structure. Additionally, we fitted a temperature gradient model to the selected wavelength range of PIONIER, GRAVITY, and MATISSE to derive the physical properties of the inner regions.}
   {
   Our parametric model determined an inclination of $\approx 47.5^\circ$ and a position angle of $\approx 83.6^\circ$, which corroborates the strong misalignment of the inner to the outer disk. From the symmetric temperature gradient, we derive an inner disk radius of $\approx0.27$~au, with dust surface densities of $\Sigma_{\rm{subl}} \approx 10^{-3.2}$ g/cm$^2$ and a vertical optical depth $\tau_{\rm{z, subl}} \approx 0.1-0.06 $.    
   Same-night MATISSE and GRAVITY observations show directional discrepancies that are inconsistent with a first-order azimuthally modulation ring. This indicates that higher-order asymmetries are required to explain the interferometric signals. This interpretation is further supported by a MATISSE snapshot image reconstruction that revealed a two-component asymmetric structure. }
 {
 The chromatic interferometric data reveal that higher-order asymmetries are probably required to explain the inner disk of HD~100453, which suggests a possible origin in dynamic interactions or disk instabilities. Coordinated multi-wavelength infrared interferometric observations with GRAVITY and MATISSE will be crucial to confirm these findings and uncover their underlying nature.}
 
   \keywords{Protoplanetary disks --
            Stars: formation --
            Methods: observational -- 
            Techniques: interferometric --
            Stars: individual: HD~100453 --
            Stars: variables: T Tauri, Herbig Ae/Be
            }

   \maketitle

\section{Introduction}
Since the first exoplanet discoveries in the early 1990s, it has been shown that many stars host planets in their inner few astronomical units (au; e.g. \citeads{1995Natur.378..355M}; \citeads{2024PSJ.....5..152L}). These close-in planets motivated the investigation of the central regions of protoplanetary disks to expand and refine planet formation theories \citepads{2023ASPC..534..717D}.

In the past decade, research with facilities such as Atacama Large Millimeter/submillimeter Array (ALMA), Gemini Planet Imager (GPI), and Spectro-Polarimetric High-contrast Exoplanet REsearch (SPHERE) has shown that unlike the classical smooth-disk paradigm, the outer regions of planet-forming disks in the millimeter- and optical wavelength regimes are highly structured and frequently show features such as gaps, rings, shadows, crescents and spirals in dust and gas (e.g. \citeads{2023ASPC..534..605B}; \citeads{2023ASPC..534..423B}). 
In parallel, high-resolution interferometric observations with the Very Large Telescope Interferometer (VLTI) have proven to be powerful tools in the infrared that can probe the inner regions of planet-forming disks at spatial resolutions of a few milliarcseconds (mas). This enables complementary and unique investigations of structures in the regions of the inner few au in the planet-forming disks (e.g. \citeads{2017A&A...599A..85L};\citeads{2019A&A...632A..53G};\citeads{2021A&A...655A..73G}).

Some circumstellar disks show narrow or broad darker regions in scattered light that are interpreted as shadow lanes and have been thought to be caused by a misalignment between the inner and outer disk due to a warp or tilt (e.g.  \citeads{2017AJ....154...33A}, \citeads{2017A&A...597A..42B}, \citeads{2023ASPC..534..605B}). The main theories for inducing such a warped or misaligned disk are a stellar or planetary mass companion on an inclined orbit or a misaligned stellar magnetic moment \citepads{2018MNRAS.473.4459F}.  
The occurrence rate of inner-outer disk misalignments is uncertain, but some statistical studies have been made. \citetads{2022A&A...658A.183B} investigated 20 transition disks with known shadow lanes with near-infrared (NIR) interferometry and reported that six of them at least showed a measurable misalignment.  \citetads{2024ApJ...961...95V} explored lateral asymmetries in edge-on disks and suggested that these phenomena might also arise from a misaligned inner disk region. Notably, 15 of the 20 investigated edge-on disks showed this lateral asymmetry. This suggests that misalignments between inner and outer disk regions are relatively common overall.

We focus on one of these objects with a significant misalignment between the inner and outer disk: the Herbig Ae system HD~100453. The star is surrounded by a transition disk with a large-scale spiral feature in its outer disk that is visible in both mm and scattered light, and a clear shadow lane that indicates disk misalignment (\citeads{2017A&A...597A..42B}; \citeads{2020MNRAS.491.1335R}). 
The large gap between the inner and outer disk is thought to be caused by an undetected exoplanet ($M_{\mathrm{p}} \leq 5 \mathrm{M_{Jup}}$) located between 15 and 20 au \citepads{2020MNRAS.499.3857N}. On the other hand, the spiral feature and the misalignment of the outer disk are attributed to the stellar companion, which orbits at a projected distance of $\sim 109$ au. This companion orbits on a significant misaligned orbit that affects the outer disk through the Kozai-Lidov mechanism (\citeads{2009ApJ...697..557C}; \citeads{2020MNRAS.499.3837G}, \citeads{2023A&A...675L...1X}).
Recent VLT/SPHERE observations ruled out inner companions with $M_{\mathrm{p}} \geq 50 \mathrm{M_{Jup}}$ in the separation range of 10 to 190 mas (1.0 to 19.7 au) from the central star \citepads{2024A&A...682A.101S}. The misaligned inner disk has also been observed with the VLTI Precision Integrated-Optics Near-infrared Imaging ExpeRiment (PIONIER) and GRAVITY instruments, which found approximate half-light radii of $2.63\pm 0.06$ mas~(0.27 au) at 1.65~$\mathrm{\mu}$m, and $3.04\pm 0.05$~mas (0.31 au) at $2.25~\rm{\mu} $m, respectively (\citeads{2017A&A...599A..85L}; \citeads{2022A&A...658A.183B}). The recovered inclination and position angle aligned well with the shadow lane in scattered light SPHERE data. The GRAVITY closure phase data showed a preference for a first-order asymmetry, with a flux skewness of A$_{\rm{skw}}\approx 14.6 \pm 1~\%$ along the major axis ($\theta_{\rm{skwPA}} \approx -106^\circ \pm 4^\circ$) in the south-west.
In Table \ref{tab:system_parameters} we summarise a few of these relevant system properties in the context of our research.  

We present the first interferometric L band observations of the inner disk region of HD~100453 with the aim to describe the continuum dust structures and asymmetry. In Sect. \ref{section:observations} we describe the observations, selection, data reduction, and calibration, and we discuss the reduced observations. In Sect. \ref{sec:results_modeling} we present the results of the parametric modelling, of the time variability analysis, and of the image reconstruction.
In Sect. \ref{sec:combined_data} we present a combined model using published PIONIER and GRAVITY data. Finally, we discuss in Sect.\ref{section:discussion} some interpretations and implications of our findings.

\begin{table}[ht]
\caption{Stellar and disk properties of the HD~100453 system.}
\label{tab:system_parameters}
{\renewcommand{\arraystretch}{1.1}
\begin{tabular}{llc}
\hline \hline
Parameter     & Value & Reference \\ \hline
              &                                 &           \\
\multicolumn{3}{l}{HD~100453 (A)} \\\hline
Spectral Type &   A9 – F0                       &   (1)        \\
Mass          &   $1.6 \pm 0.05$ M$_\odot$      &   (1)        \\
Radius          &   $1.58 \pm 0.06$ R$_\odot$      &   (1)        \\
$T_{\mathrm{eff}}$          &   $7250 \pm 125$ K              &    (1)       \\
Age           &   6.5 – 19.2 Myr                &   (1), (2)     \\
Distance      &   $103.61 \pm 0.24$ pc          &    (1)       \\
Luminosity    &   $6.2 \pm 0.14$ L$_\odot$      &   (1)        \\
              &       &           \\
\multicolumn{3}{l}{Companion (B)}  \\\hline
Spectral Type &  M4.0V – M4.5V                 &    (3)       \\
Mass          &   $0.18 \pm 0.03$ M$_\odot$      &   (3)        \\
Age           &   8 – 12 Myr               &    (3)       \\
Projected separation\tablefootmark{a}   & 1.05" ($\sim$109 au)   &     (3)      \\
              &       &           \\
              
\multicolumn{3}{l}{Disk(s)}        \\\hline
Inclination (outer) &   38$^\circ$    &  (4)          \\ 
Position angle (outer)\tablefootmark{b}&  142$^\circ$     & (4)          \\ 
Inclination (inner) &   48.7$^\circ$ – 46.1$^\circ$    &  (5), (6)          \\ 
Position angle (inner)\tablefootmark{b}&  80.8$^\circ$ – 81.6$^\circ$     & (5), (6)        \\ 
Asym. angle (inner, K band)\tablefootmark{b}&  $-106^\circ \pm 4 ^\circ$     & (6)        \\ 
  &       &           \\
  
\multicolumn{3}{l}{Hypothesised planet}            \\ \hline
Mass (theory)        &   $\leq 5 \mathrm{M_{Jup}}$      &   (7)        \\
Mass (non-detection) &  $\leq 50 \mathrm{M_{Jup}}$ & (8) \\
Estimated distance           &   15 -- 20 au               &    (7)       \\
Estimated period   & 45 -- 72 yr   &           \\
              &       &           \\ \hline
\end{tabular}
}
\tablefoottext{a}{Outside L band AT field of view, pinhole $= 1.5\lambda/D \approx$ 0.60".}
\tablefoottext{b}{All angles are defined as north to east.}

\tablebib{(1)~\citetads{2021A&A...650A.182G};
(2) \citetads{2018A&A...620A.128V}; (3) \citetads{2009ApJ...697..557C}; (4) \citetads{2017A&A...597A..42B};
(5) \citetads{2017A&A...599A..85L}; 
(6) \citetads{2022A&A...658A.183B};
(7) \citetads{2020MNRAS.499.3857N};
(8) \citetads{2024A&A...682A.101S}
}
\end{table}

\section{Observations and data reduction}
\label{section:observations}

\subsection{VLTI/MATISSE}
\begin{figure}
   \centering
   
      \includegraphics[trim={1cm 0cm 3cm 0.5cm},clip, width=8.5cm]{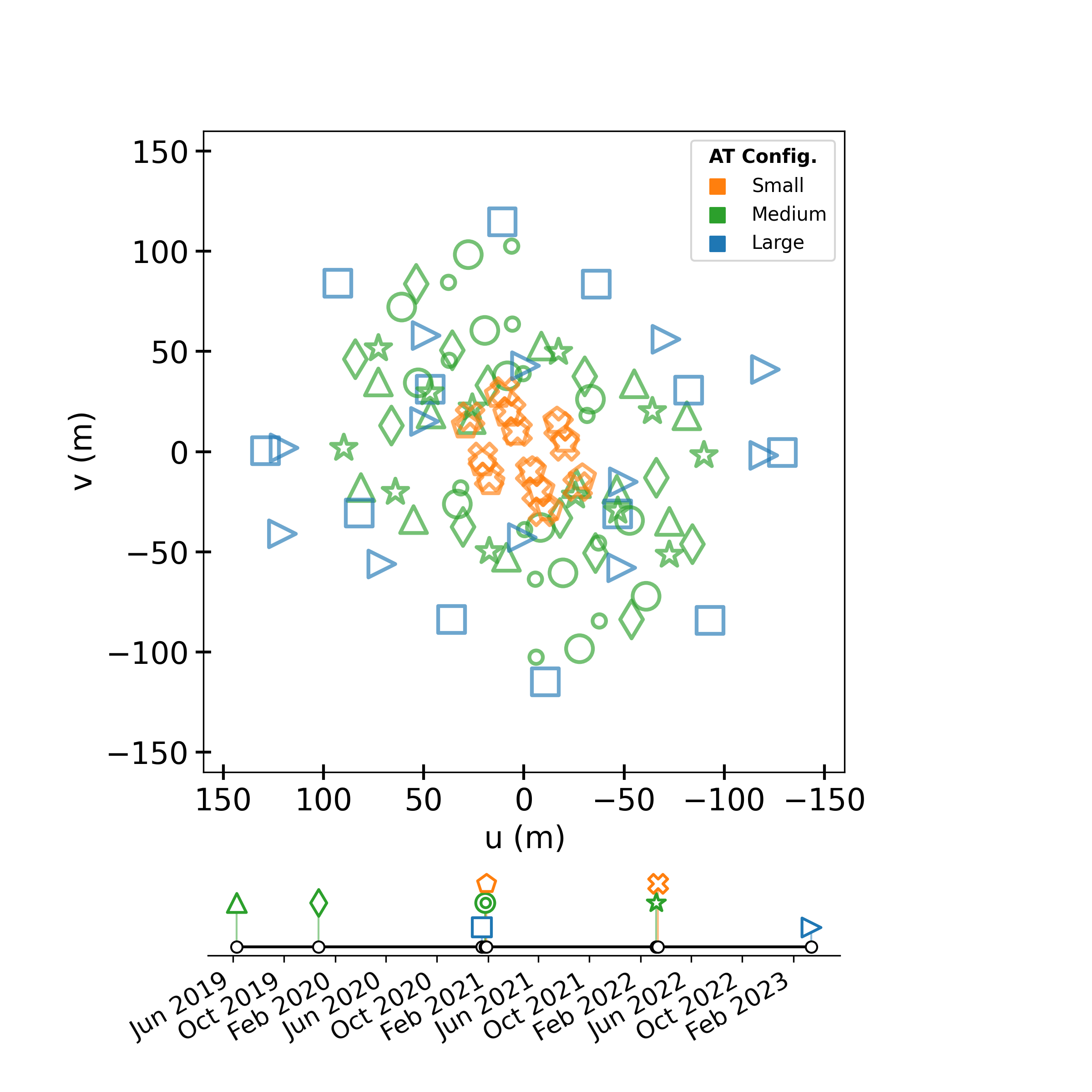}
      \caption{Overview of the $uv$ space and timeline covered by the selected {MATISSE} observations. Each night is indicated with a unique marker. The colour represents the AT station configurations. 
           }
         \label{Fig:UV}
\end{figure}

The observations of HD~100453 were made with the Multi AperTure mid-Infrared SpectroScopic Experiment (MATISSE; \citeads{2014Msngr.157....5L}; \citeads{2022A&A...659A.192L}). MATISSE can interferometrically combine the light of a target in the L, M, and N bands (3.0$-$4.0~$\mathrm{\mu}$m, 4.6$-$5.0~$\mathrm{\mu}$m, and 8$-$13~$\mathrm{\mu}$m) with either the four 8.2~m diameter Unit Telescopes (UTs) or the 1.8~m diameter Auxiliary Telescopes (ATs), with baselines spanning from $\sim$15 to $\sim$200~m. 
The data for HD~100453 were obtained with the UTs and a variety of AT arrays between May 2019  and March 2023 as part of the MATISSE guaranteed time observing campaign. Table \ref{tab:observations} shows the relevant night and observation information, and Fig.\ref{Fig:UV} shows the $(u,v)$ space sampling from the selected datasets. 
With the baselines we used, we probed the disk structures at a beam resolution of about $\theta \approx \frac{\lambda}{B} =  0.59~{\rm au} \cdot \left ( \frac{130~{\rm m}}{B} \right )\left ( \frac{\lambda }{ 3.6~{\rm \mu m}} \right )\left ( \frac{d}{103.6~{\rm pc}} \right )$. 

The typical observation sequence consisted of a science target and a calibrator star, which were both observed with two independent observation modes: chopped and non-chopped. For chopped observations, the telescope field of view alternated back and forth between taking data from the science target to taking the empty sky background, which was used to measure more accurate total fluxes. Non-chopped observations remained on target during the full duration, which enhanced the stability of the interferometric fringe tracking of the observation. This often results in better signal-to-noise ratios (S/N), but can be biased at longer wavelengths because the atmospheric conditions are different from those when the the background calibration is taken. When both were available in an observation sequence, we used the visibilities of our chopped observations and the closure phases of the non-chopped observations (for more details, see \citeads{2022A&A...659A.192L}).

All datasets were reduced and calibrated using the pipeline EsoRex DRS 2.0.2 \citepads{2016SPIE.9907E..23M} and the MATISSE pipeline tools\footnote{MATISSE pipeline tools are available separately as an addition to the standard DRS software; \url{https://github.com/Matisse-Consortium/tools}}. 
The DRS pipeline reduction followed standard settings for incoherent reduction, with two key modifications. The smoothing parameter \texttt{hampelFilterKernel} was set to 10 pixels to improve the photometric estimation (useful for faint stars $\leq5~\rm{Jy}$ in L band AT). Additionally, the sliding-average \texttt{spectralBinning} parameter was set to 5 or 75 pixels per channel for low ($\lambda / \Delta\lambda \approx 34$) or medium ($\lambda / \Delta\lambda \approx 500$) resolution, respectively, which increased the S/N and brought the observation types effectively to the same resolution.

There are clear differences in the data quality of each dataset after reduction that originate from different night conditions. 
After we assessed the data, we found that a number of our observations had to be excluded. In particular, the May 2019 UT, March 2022 ATs, and December 2022 AT sets were of poor quality. These observations were all taken during turbulent atmospheric conditions, which made the interferometric fringe-tracking and measurements less reliable. 
The faintness of HD~100453 ($\leq 4~\rm{Jy}$ in correlated fluxes) in the N band is below the recommended AT sensitivity limit, and the data are therefore very noisy at longer AT baselines. The weather conditions during the May 2019 UT observations also resulted in poor N-band data, although this might be salvageable in the future by a planned improvement in the frame-flagging algorithm in the data reduction pipeline. We therefore limited our study to the L band continuum. 

Unfortunately, we were not able to always use the entire spectral coverage offered by the MATISSE L band because frequent artefacts arose from atmospheric H$_{2}$O absorption. Consequently, the starting segment of the L band ($ < 3.4~\mu$m) had to be excluded from all datasets. Additionally, visibilities at longer wavelengths in especially non-chopped observations were frequently biased by the background subtraction, and they were therefore also removed.  The upper limits were chosen manually per dataset in the range of $3.6 - 3.8 ~\mu$m, depending on where the distances between the reduced visibilities stayed roughly constant.
As a result, our analysis primarily focuses on the continuum emission from 3.4 to $3.8~\mu$m (see Appendix \ref{appendix:observations}).

After the reduction and calibration, each measurement provided us with the total flux and the following interferometric quantities: 
six squared visibilities \mbox{($V^2_{ij}(\lambda)= |I_{ij}(\lambda)|^2 / (P_{i}(\lambda) \cdot P_{j}(\lambda*$)}, 
and four closure phases ($\Phi_{ijk}(\lambda) = Arg [ I_{ij}(\lambda) \cdot I_{jk}(\lambda) \cdot I_{ik}^*(\lambda) ] $), where $I(\lambda)$ is the
Fourier transform of the interferogram, $P(\lambda)$ is the photometric flux, and $V^2_{ij}(\lambda)$ is the normalised squared visibility amplitude between a telescope pair $i$ and $j$. The indices $i$, $j$, and $k$ correspond to individual telescopes.  

\begin{figure*}
   \centering
   \includegraphics[width=19cm]{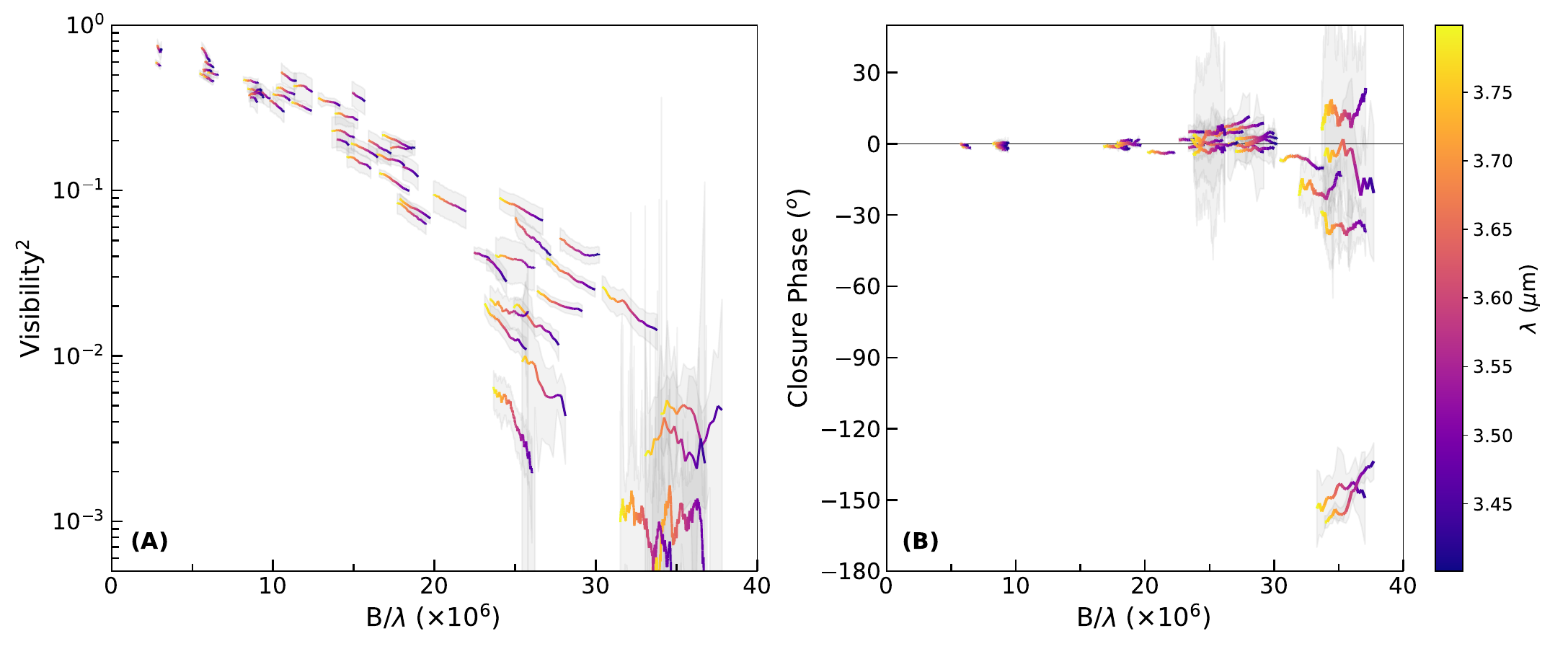}
      \caption{Calibrated squared visibilities and closure phases from all the selected MATISSE observations.
           }
         \label{Fig:overview_data_all}
\end{figure*}

Figure \ref{Fig:overview_data_all} shows an overview of the visibilities $V(\lambda)$ and closure phases $\Phi(\lambda)$ of the MATISSE L band data that combines all observing dates and only includes the selected wavelengths (see Appendix \ref{appendix:observations} for the results for each individual observing date). The visibilities in Fig. \hyperref[Fig:overview_data_all]{2a} decrease from $\sim$0.9 on short baselines ($\sim$3 M$\lambda$) to $10^{-3}$ on baselines $>$30 M$\lambda$. 
The visibilities do not reach 1.0 on short baselines, which indicates that $\sim$10\% of the flux is distributed on scales $>$35 mas ($<$3 M$\lambda$).  
Observations on baselines with similar lengths, but different position angles show a different slope of the visibility amplitude versus baseline curve. This indicates that the emission follows an elongated shape.
Figure \hyperref[Fig:overview_data_all]{2b} shows that the closure phases reach a few degrees out to 20 M$\lambda$ and start to deviate by 10$^\circ$s to 150$^\circ$ on longer baselines. The latter suggests an asymmetry on scales of $\sim$3 mas. 

\subsection{VLTI/PIONIER and VLTI/GRAVITY archival data}
To complement our modelling, we also included published calibrated interferometric data in H band ($1.5 - 1.8~\mu$m) and K band ($2 - 2.5~\mu$m) from the PIONIER and GRAVITY VLTI instruments, respectively. The data from programme IDs  190.C-0963 \& 106.21JR.001 were retrieved from the JMMC Optical interferometry DataBase\footnote{\url{http://oidb.jmmc.fr/}} or were directly provided by the principal investigators. The 11 PIONIER observations cover earlier times between December 2012 and February 2013, and the 7 GRAVITY datasets fall within our MATISSE timeline, with data between December 2020 and March 2021 (for details of the reduction and observation for the PIONIER and GRAVITY data, see \citeads{2017A&A...599A..85L} and \citeads{2022A&A...658A.183B}).\\

\section{Analysis of MATISSE observations} 
\label{sec:results_modeling}
In this section, we present the analysis of the MATISSE data in three parts. In Sect. \ref{subsec:matfirst} we describe our parametric time-invariant chromatic modelling, and in Sect. \ref{subsec:time_variability} we examine the time variability of the observed asymmetric dust feature, followed by an exploration of the image reconstruction in Sect. \ref{subsec:reconstruction}.

\subsection{Parametric modelling}
\label{subsec:matfirst}
For the parametric modelling of the squared visibilities $V^2_\lambda$ and closure phases $\Phi_\lambda$,
we employed the open-source software package \texttt{oimodeler}\footnote{\href{https://github.com/oimodeler/oimodeler}{https://github.com/oimodeler/}.} \citepads{2024SPIE13095E..2WM}. Following the method of \citeads{2017A&A...599A..85L}, we assumed a disk model constructed with an asymmetric infinitesimally thin ring, described with parameters $r_{\rm{ring}}, \theta_{\rm{PA}}$ and $e$, plus a first-order azimuthal modulation, described with parameters $A_{\rm{skw}}$ and $\theta_{\rm{skwPA}}$. The complex visibilities $V_{\rm thin\; ring}$ are given by

\begin{align}
    V_{\rm{thin \; ring}}(q) &= J_0(2\pi r_{\rm{ring}} q)
    - i J_1(2\pi r_{\rm{ring}}q) \cdot A_{\rm{skw}} \sin(\Psi),  \\
    \rm{with}\nonumber\\
    \Psi &= \theta_{u,v} + (\theta_{\rm{skwPA}} - \theta_{\rm{PA}}) \nonumber \\
    q &= \bigl[ \left ( \left (u \cos(\theta_{\rm{PA}})-v \sin(\theta_{\rm{PA}})  \right ) / e  \right )^{2} + \nonumber \\ 
               & \;\;\;\;\;\left (u \sin(\theta_{\rm{PA}})+v \cos(\theta_{\rm{PA}})  \right )^{2} \bigr] ^{0.5} \nonumber.
\end{align}

Here, $J_0$ and $J_1$ are the Bessel functions. This infinitesimally thin modulated ring was convolved with a pseudo-Lorentzian/Gaussian kernel with parameters $r_{\rm{k}}$ and $f_{\rm{Lor}}$ to describe the radial extent of the ring. This resulted in the following description for the complex visibility:

\begin{align}
    V_{\rm{kernel}}(q) &= (1-f_{\rm{Lor}})\cdot\exp(-\frac{(\pi r_{\rm{k}} q)^2 }{\ln(2)})  + f_{\rm{Lor}} \cdot \exp(-\frac{2\pi r_{\rm{k}}q  }{\sqrt{3}} ) \nonumber \\
    V_{\rm{asym. \; ring}}(q) &= V_{\rm{thin \; ring}}(q) \cdot V_{\rm{kernel}}(q).
\end{align}

\noindent Furthermore, the stellar contribution was assumed to be unresolved $V(q) = 1$. 
The flux contributions are chromatic and follow a spectral power law, that is, for the circumstellar material, $V_{\rm{asym. \; ring}}(q)*f_c (\frac{\lambda}{\lambda_0})^{k_c}$, similar to Eq. (4) in \citeads{2017A&A...599A..85L} ($\lambda_0 = 3.5  \mu$m). The stellar spectral index was determined by its blackbody temperature, and the extended emission was assumed to follow the same power law as the circumstellar material. 
We also tested a varying spectral index of the extended halo, but this had little effect on the fit quality. The parameters defining this model are further listed in Table \ref{tab:model_parameters}.

\begin{table}[ht]
\caption{Description of the model parameters.}
\label{tab:model_parameters}
{\renewcommand{\arraystretch}{1.1}
\begin{tabular}{l p{65mm}} 
\hline \hline
Parameters & Description  \\ \hline 
$l_{\rm{a}}$          & 10-log of the approximate half-light radius: $l_{\rm{a}} = \log_{10}(\sqrt{r_{\rm{ring}}^2 + r_{\rm{k}}^2})     $             \\ 
$l_{\rm{kr}}$          & 10-log of the kernel half-light ratio: $l_{\rm{kr}} = \log_{10}(r_{\rm{k}} / r_{\rm{ring}})$                                \\
$f_{\rm{Lor}}$         & Fraction of the pseudo-Lorentzian versus Gaussian profile in the kernel. \\
$\theta_{\rm{PA}}$         & Position angle$^a$ of the ellipse semi-major axis.     \\
$e$       & Ring elongation = $1 / \cos(\phi_{\rm{inclination}})$.                         \\
$A_{\rm{skw}}$        & Amplitude of the azimuthal modulation.                  \\
$\theta_{\rm{skwPA}}$      & Position angle$^a$ of the asymmetry. \\ 
$f_{\rm{s}}$      & Fraction of the total flux originating from the unresolved central star. 
\\ 
$f_{\rm{c}}$      & Fraction of the total flux originating from the circumstellar material. 
\\ 
$f_{\rm{h}}$      & Fraction of the total flux completely resolved at the shortest baselines of our data, representing a spatially extended emission component (also known as `halo'), fixed as $f_{\rm{h}}$ $= 1 - f_{\rm{s}} - f_{\rm{c}}$.\\  
${k_c}$ & Spectral index of the circumstellar material: $f_{\rm{c}}(\frac{\lambda}{\lambda_0})^{k_c}$.\\ \hline
\end{tabular}
}
\tablefoottext{a}{Defined as counter-clockwise, i.e. N$\rightarrow$E $=0^\circ\rightarrow90^\circ$.}\\
\end{table} 

Table \ref{tab:model_fit_onlymatisse} shows the best-fit parametric values found by a $\chi^2$-minimisation with the Markov chain Monte Carlo (MCMC) algorithm using the \texttt{emcee} Python package \citepads{2013PASP..125..306F}. For the parametric models, we used 32 walkers, 4000 steps, 1500 steps of burn-in time, and flat priors. We explored three model variants: model (1) was a symmetric inclined dust ring, with a stellar flux contribution and larger-scale flux captured as `halo'; model (2) was the same, but a  first-order azimuthal modulation was added to the ring; model (3) was the same as model (2), but the disk orientation, size, and flux ratio were fixed to the best-fit values for model (1). 

The models yield an inclination of $47.5 - 48.2$~degrees and a position angle of $83.6 -85.9$~degrees. This agrees to within a few degrees with previous interferometric work in the H and K bands, as shown in Table \ref{tab:system_parameters}.
 The half-light radius size, $\sqrt{r_{\rm{ring}}^2  + r_{\rm{k}}^2} \approx 3.89 $ mas, is slightly larger than for GRAVITY ($\approx 3.02 $ mas) and PIONIER ($\approx 2.63$ mas; \citeads{2017A&A...599A..85L}; \citeads{2022A&A...658A.183B}), which shows a modest but clear increase with wavelength. This is expected because at longer wavelengths, the radiation contribution of cooler material that is located farther away from the star increases.
 
\begin{figure*}
   \centering
   \includegraphics[width=19cm]{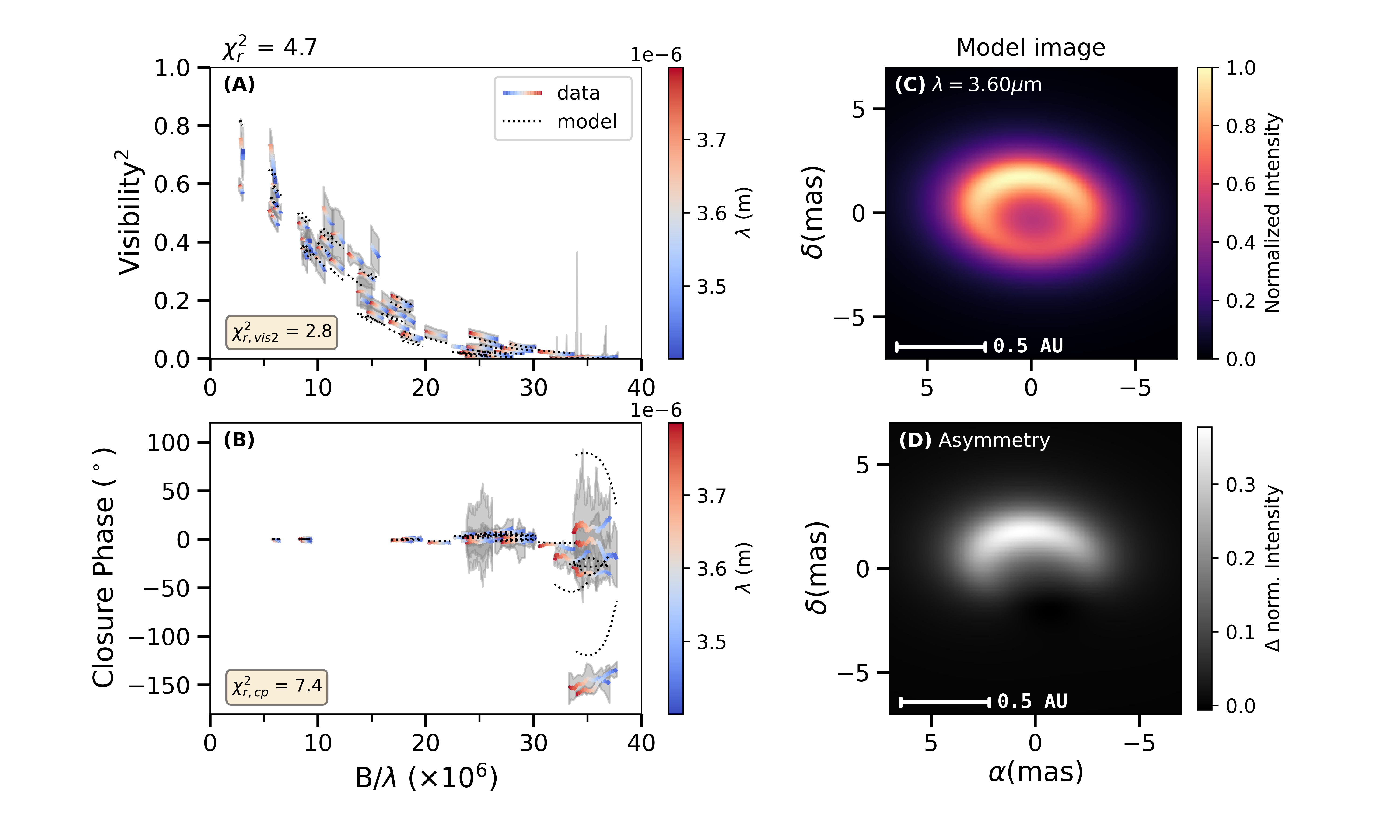}
      \caption{Fit of the MATISSE data following a similar time-invariant parametric model and method as \citeads{2017A&A...599A..85L} and \citeads{2022A&A...658A.183B} (model 2; see Table \ref{tab:model_fit_onlymatisse}).  Panels A and B: Observations and the best-fit model simulated squared visibilities and closure phases. Panel C: Model brightness distribution $I_\nu\rm{(fit)}$.\ Panel D: Skewed asymmetry following Eq. \ref{eq:highlight_asymm}.
           }
         \label{Fig:model_only_matisse_data}
\end{figure*}

The data also show that an asymmetry needs to be included in the disk because of the significant closure phases on the longest baselines. 
Figure \ref{Fig:model_only_matisse_data} shows the best-fit model, which corresponds to an stationary asymmetric ring model. 
Table \ref{tab:model_fit_onlymatisse} shows that models (2) and (3) differ slightly in the strength and position of the asymmetry, depending on the order in which we fitted the disk parameters.
Nevertheless, they still agree overall for the asymmetry with stronger emission north-east of the disk.

\begin{table}[ht]
\caption{Fit results of the parametric models to the L band data. }
\label{tab:model_fit_onlymatisse}
{\renewcommand{\arraystretch}{1.4}
\begin{tabular}{l l p{18mm} p{18mm} p{18mm}} 
\hline \hline
\multicolumn{2}{l}{Parameters} &  (1) Sym. & (2) Asym. & (3) Asym.  \\ \hline
$l_{\rm{a}}$             & (mas)      & $0.590_{-0.001}^{+0.001}$   & $0.589_{-0.001}^{+0.001}$ & $^{(1)}$  \\
$l_{\rm{kr}}$            &      & $-0.041_{-0.002}^{+0.002}$ &$-0.005_{-0.003}^{+0.002}$ & $^{(1)}$  \\
$f_{\rm{Lor}}$           &          & $1.000_{-0.001}^{+0.}$       &$0.960_{-0.004}^{+0.005}$ & $^{(1)}$  \\
$e$                      &            & $1.500_{-0.002}^{+0.002}$ &$1.481_{-0.002}^{+0.002}$ & $^{(1)}$  \\
($\theta_{\rm{inc}}$)\tablefootmark{a} & ($^\circ$) &($48.19_{-0.07}^{+0.07}$) & ($47.53_{-0.07}^{+0.07}$) & $^{(1)}$  \\
$\theta_{\rm{PA}}$       & ($^\circ$) & $85.88_{-0.17}^{+0.15}$   &$83.57_{-0.14}^{+0.14}$ & $^{(1)}$  \\
$A_{\rm{skw}}$          &            & -                         & $ 0.315_{-0.003}^{+0.003}$ & $0.282_{-0.003}^{+0.003}$ \\
$\theta_{\rm{skwPA}}$   & ($^\circ$) & -                         & $18.33_{-0.18}^{+0.17}$ & $14.54_{-0.03}^{+0.03}$ \\ 
$k_c$                    &            & $0.37_{-0.06}^{+0.07}$    &$1.83_{-0.10}^{+0.10}$ & $^{(1)}$  \\
$f_{\rm{s}}$ & & $0.040_{-0.012}^{+0.007}$ &$0.031_{-0.009}^{+0.006}$ & $^{(1)}$  \\
$f_{\rm{c}}$ & & $0.735_{-0.222}^{+0.136}$ &$0.75_{-0.21}^{+0.14}$ & $^{(1)}$  \\ \hline
$\chi^2_{\rm{r, vis^2}}$          & & 2.8   & 2.8  & 2.8 \\
$\chi^2_{\rm{r,} \Phi_{\rm{cp}}}$ & & 13.4 & 7.4 & 7.4 \\
$\chi^2_{\rm{r}}$                 & & 7.0 & 4.7 & 4.7 \\ \hline
\end{tabular}
}
\tablefoot{The values represent median values with 16\%--84\% quantile values. Models 1, 2, and 3 represent a symmetric ring, an asymmetric ring, and an asymmetric ring. All parameters except those controlling the asymmetry were fixed to the best-fit values of the symmetric model (see Appendix \ref{appendix:error} for a discussion of the uncertainties).\\
\tablefoottext{a}{Elongation was converted to inclination for convenience.}\\
\tablefoottext{1}{Same value as model 1.}
}
\end{table}

\subsection{Constraints on the time variability through MATISSE epochs}
\label{subsec:time_variability}
Our parametric fit places the asymmetry at an approximate projected distance of $\approx$0.4~au with respect to the central star.
Assuming Keplerian orbital motion, the period of an object at this distance is $\approx 73.9 \pm 1.4$ days (using the mass and distance from Table \ref{tab:system_parameters}). This raises the question whether the disk parameters obtained from the asymmetric-ring fit of Sect. \ref{subsec:matfirst} are affected by smearing in the time domain. To investigate the variability of the asymmetry, we split the dataset into six separate time epochs (see Table \ref{tab:observations}), each containing at least a medium or large AT array. For each epoch, we fitted the asymmetry amplitude $A_{skw}$ and angle $\theta_{skwPA}$ and kept the other disk parameters fixed to the best-fit values of asymmetric model (2). 

\begin{figure}
   \centering
   \includegraphics[width=9cm]{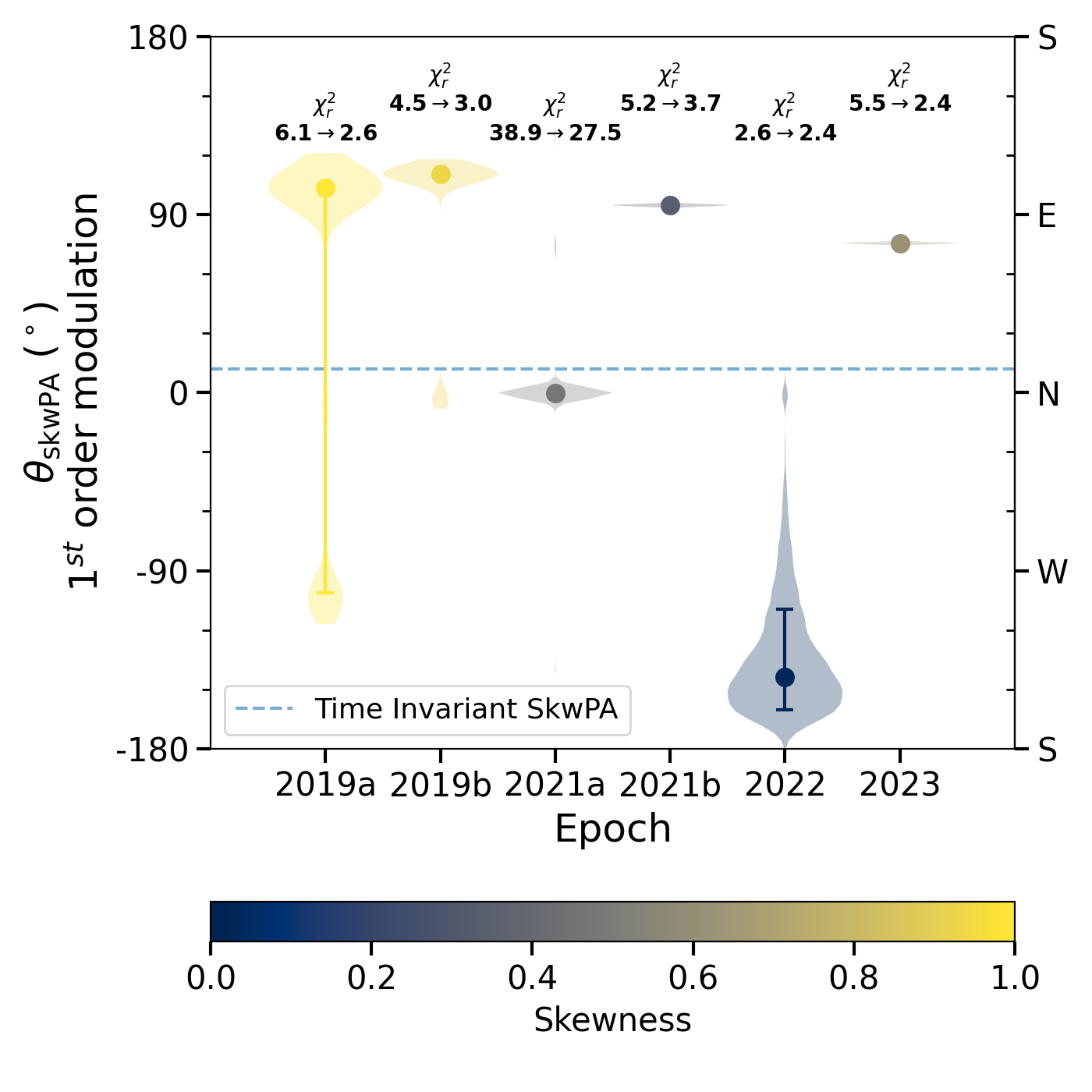}
      \caption{Violin-plot of the time-variable MCMC fitting results of the asymmetry angle ($\theta_{\rm{skwPA}}$) for six epochs of MATISSE data. The other disk parameters were fixed to the best-fit model 2 asymmetric disk parameters from Table \ref{tab:model_fit_onlymatisse}. The horizontal dashed line shows the best-fit value for the asymmetry angle for the time-invariant dataset. Above each violin, the change in the respective $\chi^2_r$ is shown compared to the time-invariant value. The error bars indicate the best-fit value with 16/84\% percentile intervals.  
           }
         \label{Fig:Time_Variability}
\end{figure}

The fitting results in Fig. \ref{Fig:Time_Variability} clearly show improvements to the reduced $\chi^2_r$ by making the asymmetric angle and strength epoch dependent. 
The best-fit asymmetry angle constraints differ for the epochs, likely due to variations in the number of datasets, data quality, and AT configurations.
 The non-overlapping confidence intervals between some of the epochs are a potential indication for time variability. The relatively high $\chi^2_r$ values suggest, however, that the model might still not capture the full underlying complexity.

\subsection{Image reconstruction}
\label{subsec:reconstruction}

Since the disk structures might be variable in time, we had to ensure that we did not smear out our signal by combining different epochs. Of all our epochs, the four 2021 observations were closest in time ($\sim 11$ days), and provide a well-sampled $uv$ plane.
This allowed us to perform a simplistic image reconstruction by assuming that the probed disk structures are roughly stationary over this time.

For our image reconstruction, we used the software SPARCO/MIRA v2.3.2 through the JMMC OI-imaging tool\footnote{\url{https://www.jmmc.fr/tools/data-analysis/oimaging/}} (\citeads{2008SPIE.7013E..1IT}; \citeads{2014A&A...564A..80K}). SPARCO is a parametric approach for chromatic image reconstruction, where the image is modelled as a sum of (chromatic) geometrical models and the reconstructed image. We modelled the star geometrically as a central blackbody point source with a temperature of $T = 7250$ K. For the reconstructed image chromaticity, we assumed that the disk was well described by a single-temperature blackbody at $T \approx 1445$ K, where the latter provides a good fit to the SED for our wavelength range (see App. \ref{appendix:sed}). We reiterate that this temperature holds no more physical meaning than a power law, for example, and that this method was purely used as an approximation of the wavelength dependence of the flux ratio of the inner disk relative to the star. 
We used a compactness regularisation function that favoured image structures in the central region of FWHM $\sim$40 mas. As an initial image prior, we took two circular Gaussians with FWHM $= 5.5$ and 40 mas, with a total field of view of 70$\times$70 mas and a size of 0.5 mas per pixel.
We ran the algorithm twice consecutively with 200 iterations and took the first intermediate result as the next image prior. We reached a reduced $\chi^2 \approx 0.88$.

\begin{figure*}
   \centering
    \includegraphics[trim={1cm 0cm 0cm 0cm},clip, width=17cm]{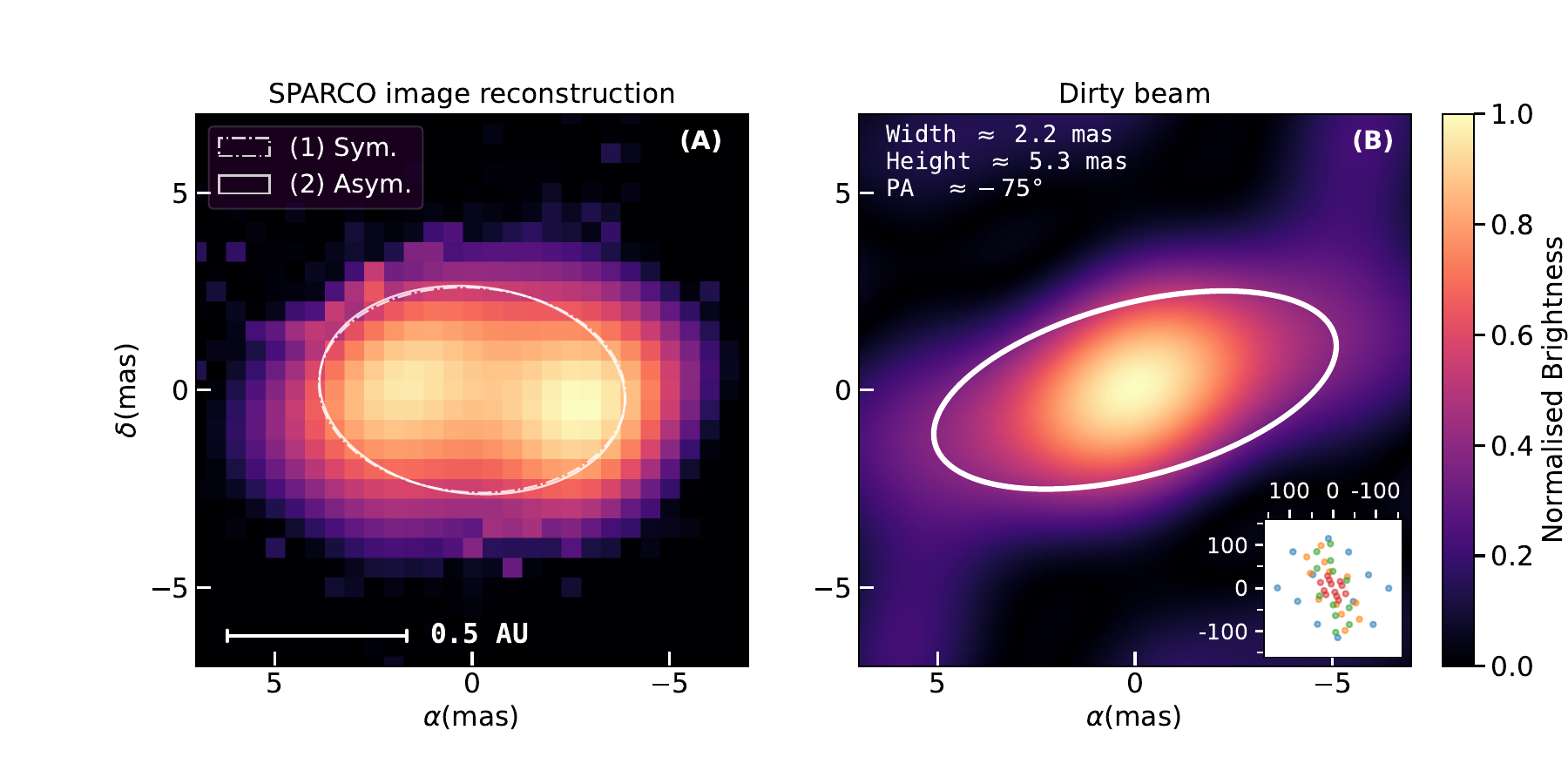}
      \caption{Panel A: L band image reconstruction for the 2021 MATISSE AT observations, which contain the large-medium-small arrays observed within 11 days, at $\lambda = 3.6 \pm 0.2~\mu$m using the SPARCO/MIRA software. The central star with $T=7250$~K, modelled as a point source centred on (0,0), is not shown. We over-plot ellipses with semi-major axes of the half-light radius $l_a$, taken from the best parametric fits of Table \ref{tab:model_fit_onlymatisse}.  Panel B: Approximate dirty beam of the combined observations at $\lambda = 3.6~\mu$m, together with a small inset of the $uv$ points.}
      \label{Fig:reconstruction} 
\end{figure*}

Figure \ref{Fig:reconstruction} shows the reconstructed MATISSE image together with the beam. The image only shows the reconstructed circumstellar emission and omits the central stellar point source. The $uv$ sampling resulted in an elliptical dirty beam with half-widths at half-maximum of $2.2-5.3$ mas and weak side lobes. 
On the large scale, we found no significant L band structure, except for an imprint of the side lobes. This is expected because we lack the fidelity to detect the $\approx10\%$ of flux that is distributed over this larger area. 
On smaller scales, the general emission agrees with the position angle and shape found in the parametric modelling. Moreover, to explain the closure phases, the algorithm preferred to place some emission in two separate components approximately east and west of the star. We experimented with priors and the number of iterations for the algorithm and found that it had no significant effect on the resulting position of the main bright structures, but instead altered the details of the shapes or gradually start to overfit.

\section{Multi-instrument analysis}
\label{sec:combined_data}

To assess whether a morphology with a more physically temperature structure can reproduce the observed data, we integrated our MATISSE observations with prior PIONIER and GRAVITY results. This combined dataset spans a broad wavelength range of 1.5 - 3.8~$\mu$m. Using this coverage, we constructed a symmetric two-dimensional (2D) power-law temperature gradient model, which also allowed us to estimate key parameters such as the dust optical depth and surface density. We computed dust opacities ($\kappa_\nu$) at the boundaries and central wavelengths of each spectral band using \texttt{optool} \citepads{2021ascl.soft04010D} assuming standard DIANA conditions; pyroxene (70\% Mg) and carbon, a dust population $n_{\rm{dust}} \propto a^{-3.5}$ with particle radius $a_{\rm{dust}} \in [0.05~{\rm \mu m}, 3 {\rm mm}]$, and a porosity of 25\% \citepads{2016A&A...586A.103W}. 

\begin{figure*}
   \centering
   \includegraphics[width=19cm]{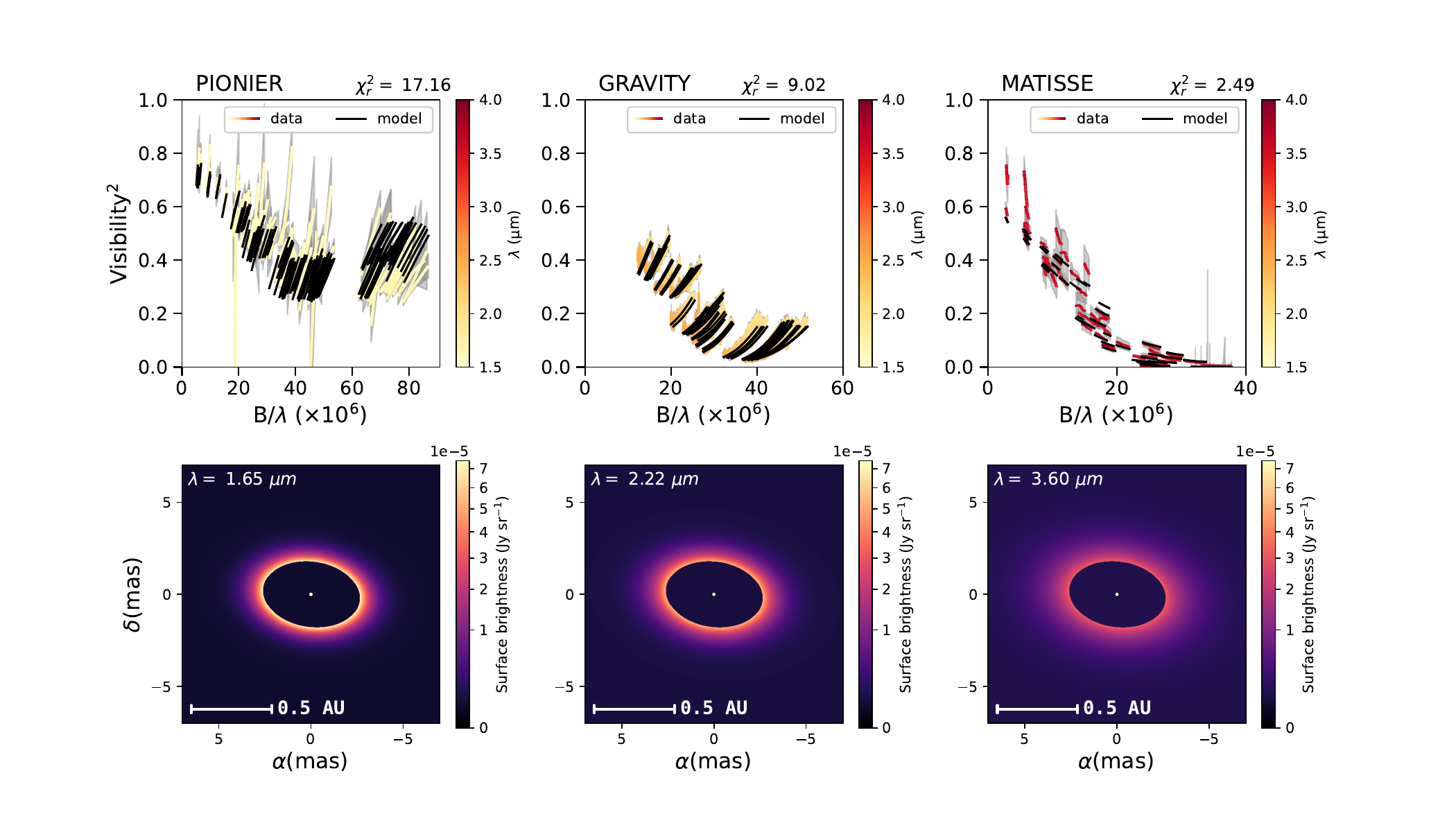}
      \caption{Best-fit symmetric temperature gradient model. Top panels: Measured and modelled visibilities. Bottom panels: Surface brightness distribution. The size of the central point source is not to scale, and the colour scaling is homogeneous for all wavelengths.
              }
         \label{Figtempgrad_model}
\end{figure*}

For our simplified model, we assumed an inclined continuous disk with temperature and surface density power-law profiles of $T(r) = T_0 (\frac{r}{R_{\rm{in}}})^{-q_T}$ and  $\Sigma(r) = \Sigma_0 (\frac{r}{R_{\rm{in}}})^{-p_\Sigma}$, which dropped to zero outside the inner and outer radii, $R_{\rm{in}}$ and $R_{\rm{out}}$, respectively. The surface brightness was calculated with the equation of radiative transfer $I_\nu(r)=(1-e^{-\tau_\nu(r)})B_\nu(T(r))$, with $\tau_\nu(r) = \Sigma(r) \kappa_\nu$, where $B_\nu$ is the blackbody radiation \citepads{1979rpa..book.....R}.
For the central point-source star, we adopted the same literature values as in previous sections (see Table \ref{tab:system_parameters}). The extended emission was modelled with a series of power laws $F(\lambda) = A_{\rm{bg}} (\frac{\lambda}{\lambda_0})^{-p_{\rm{bg}}}$ in which each instrument had its instrument-specific parameters so that it was agnostic about the overall spectral shape.
We assumed an inner dust sublimation temperature $T_0 = 1500$~K.
The position angle and inclination were fixed on the PIONIER/GRAVITY/MATISSE averaged values of the best visibility fit parametric ring models $ \theta_{\rm{PA}} \approx 82.0 ^\circ$, $\theta_{\rm{inc}} \approx 47.4 ^\circ$, which also corresponds well with the observed shadow lane in scattered light data \citepads{2022A&A...658A.183B}. The outer radius in particular is hard to constrain with our data because our infrared wavelengths mostly contain information on the warmer inner regions of the disk. We used 25 walkers, 3000 steps, 1000 steps of burn-in time, and flat priors.
Table \ref{tab:model_temperature_gradient} presents the fit parameters obtained from the MCMC analysis. To show the effect on our results, we selected two values for the outer radius, approximately 2~au and 10~au. The latter is roughly halfway to the gap edge of the outer disk as seen by ALMA. 
The two $R_{\rm{out}}$ choices yielded only marginally different values overall. Figure \ref{Figtempgrad_model} shows the best-fit model for the $R_{\rm{out}} = 2$ au case.  The inner radius falls around $0.272$ au or 2.62 mas, and additionally, the fitted inner disk surface density is approximately $10^{-3.2}$ g/cm$^2$, corresponding to optically thin vertical optical depths $\tau_{\rm z, subl} \approx 0.1-0.06 $ for the H, K, and L band wavelengths. We note that the wavelength dependence of the background halo, which is thought to be caused by quantum-heated particles \citepads{2017A&A...599A..80K}, follows a spectral shape similar to the single-temperature blackbody estimate for the SED (Fig. \ref{FigStellarFlux2}). The model underestimates the total NIR flux overall compared to the literature SED by 10 to 45$\%$ in the respective bands.

\begin{table}[ht]
\caption{Temperature gradient model-fit parameters. }
\label{tab:model_temperature_gradient}
{\renewcommand{\arraystretch}{1.4}

\begin{tabular}{l l p{1.5cm} p{1.5cm}} 
\hline \hline
Fixed Parameters & & Value & \\ \hline
$T_0$                       & (K)            & \multicolumn{2}{c}{1500} \\
PA                          & ($^\circ$)     & \multicolumn{2}{c}{82.0}  \\
incl.                       &($^\circ$)      & \multicolumn{2}{c}{47.4} \\
$R_{\rm{out}}$              & (au)           & $\;\;$2 & $\;\;$10 \\ \hline 
Parameters  & & Value & \\ \hline
$R_{\rm{in}}$                  & (au)          & $\;\;0.272^{+0.001}_{-0.001}$ & $\;\;0.272^{+0.001}_{-0.001}$  \\
                               & (mas)         & $\;\;2.63^{+0.002}_{-0.002}$   &  $\;\;2.62^{+0.002}_{-0.002}$   \\

q$_{\rm{T}}$                        &               & $\;\;1.056^{+0.002}_{-0.002}$    & $\;\;1.067^{+0.02}_{-0.02}$      \\
$\log_{10}(\Sigma_0)$          & (g / cm$^2$)  & $-3.227^{+0.001}_{-0.001}$    & $-3.208^{+0.001}_{-0.001}$      \\
$\log_{10}$(p$_{\Sigma})$     &               & $-4.36^{+0.39}_{-0.54}$    &  $-4.32^{+0.37}_{-0.54}$      \\
A$_{\rm{pio}}$ ($1.65~\rm{\mu m}$)    & (Jy)          & $\;\;0.279^{+0.002}_{-0.002}$ & $\;\;0.285^{+0.002}_{-0.002}$  \\
p$_{\rm{pio}}$                       &               & $-3.22^{+0.14}_{-0.14}$ & $-3.14^{+0.14}_{-0.13}$     \\
A$_{\rm{grav}}$ ($2.22~\rm{\mu m}$)    & (Jy)          & $\;\;0.486^{+0.001}_{-0.001}$ & $\;\;0.488^{+0.001}_{-0.001}$  \\
p$_{\rm{grav}}$                       &               & $-3.073^{+0.005}_{-0.005}$ & $-3.025^{+0.005}_{-0.005}$     \\
A$_{\rm{mat}}$ ($3.6~\rm{\mu m}$)    & (Jy)          & $\;\;0.694^{+0.002}_{-0.002}$ & $\;\;0.70^{+0.002}_{-0.002}$  \\
p$_{\rm{mat}}$                       &               & $\;\;0.47^{+0.09}_{-0.09}$ & $\;\;0.43^{+0.09}_{-0.09}$  \\
\hline
$\chi^2_{r, total}$        & & $\;\;$7.7  & $\;\;$7.5  \\
$\chi^2_{r, \rm{PIONIER}}$ & & $\;\;$16.8   & $\;\;$17.2   \\ 
$\chi^2_{r, \rm{GRAVITY}}$ & & $\;\;$9.2 & $\;\;$9.0 \\
$\chi^2_{r, \rm{MATISSE}}$ & & $\;\;$2.6   & $\;\;$2.5  \\

\hline
\end{tabular}
\tablefoot{ The fitted values represent the MCMC median with 16\% - 84\% quantile values. The reduced $\chi^2$ was calculated for the visibility data. The two columns highlight the (lacking) effect on the results by the choice for outer radius 
(see Appendix \ref{appendix:error} for a discussion of the uncertainties).
}

}
\end{table}

\section{Discussion }
\label{section:discussion}

\subsection{L band emission region and the dust sublimation radius}
Determining the temperature and structure of the inner disk rim is a complex problem that strongly depends on the dust properties.
Our best-fit inner radius from the temperature gradient model (Table \ref{tab:model_temperature_gradient}) agrees well with the evaporation radius estimated by \citeads{2017A&A...599A..80K} ($R_{\rm{evap.}} = 0.27$ au).
The temperature gradient model, however, which is based on standard DIANA conditions and was used in both studies, includes small silicate grains. This is likely inconsistent with the N-band spectra.
Previous SED analyses of HD~100453 attributed the absence of  its $10 \mu$m silicate emission feature to a depletion of small ($ \lesssim 4\mu$m) silicate grains \citepads{2003A&A...402..767M}. 
As this issue is less relevant for the H, K, and L bands because the opacity behaves more linearly, we proceeded without accounting for it here. 
Nonetheless, future high-quality N-band observations with the UT telescopes would be invaluable for constraining the dust masses, composition, grain sizes, and crystallinity in this system, similar to the approach taken by \citeads{2024A&A...681A..47V}.

\subsection{Time variability}
A known source of asymmetry at short NIR wavelengths in inclined disks is obscuration by the dust torus (\citeads{1998A&A...337..832K};\citeads{2010ARA&A..48..205D}), which causes an emission peak along the minor axis of the projected inner disk. Some possible sources of asymmetries around the major axis might be a rotating vortex, spiral, or dust clump, similar to what was observed in recent inner disk studies (\citeads{2021A&A...654A..97G};\citeads{2021A&A...647A..56V};\citeads{2023ApJ...947...68I}; \citeads{2024A&A...684A.200G}), or an eccentric disk \citepads{2013A&A...553L...3A}.

As mentioned in the introduction, \citeads{2022A&A...658A.183B} found that the HD~100453 inner disk K band emission for a time-independent fit was well described by a disk with a slight asymmetry along the major axis in the west, at $\theta_{\rm{skwPA}} \approx -106^\circ \pm 4$.
In Sect. \ref{subsec:time_variability} we determined the best-fit values for the asymmetry angle for our separate MATISSE epochs. 
Most L band epochs also favour an asymmetry that is oriented along the major axis, typically toward the east. Epoch 2021a is an exception and instead prefers an orientation along the northern minor axis (see Fig. \ref{Fig:Time_Variability}).

\begin{figure}
   \centering
   \includegraphics[width=9cm]{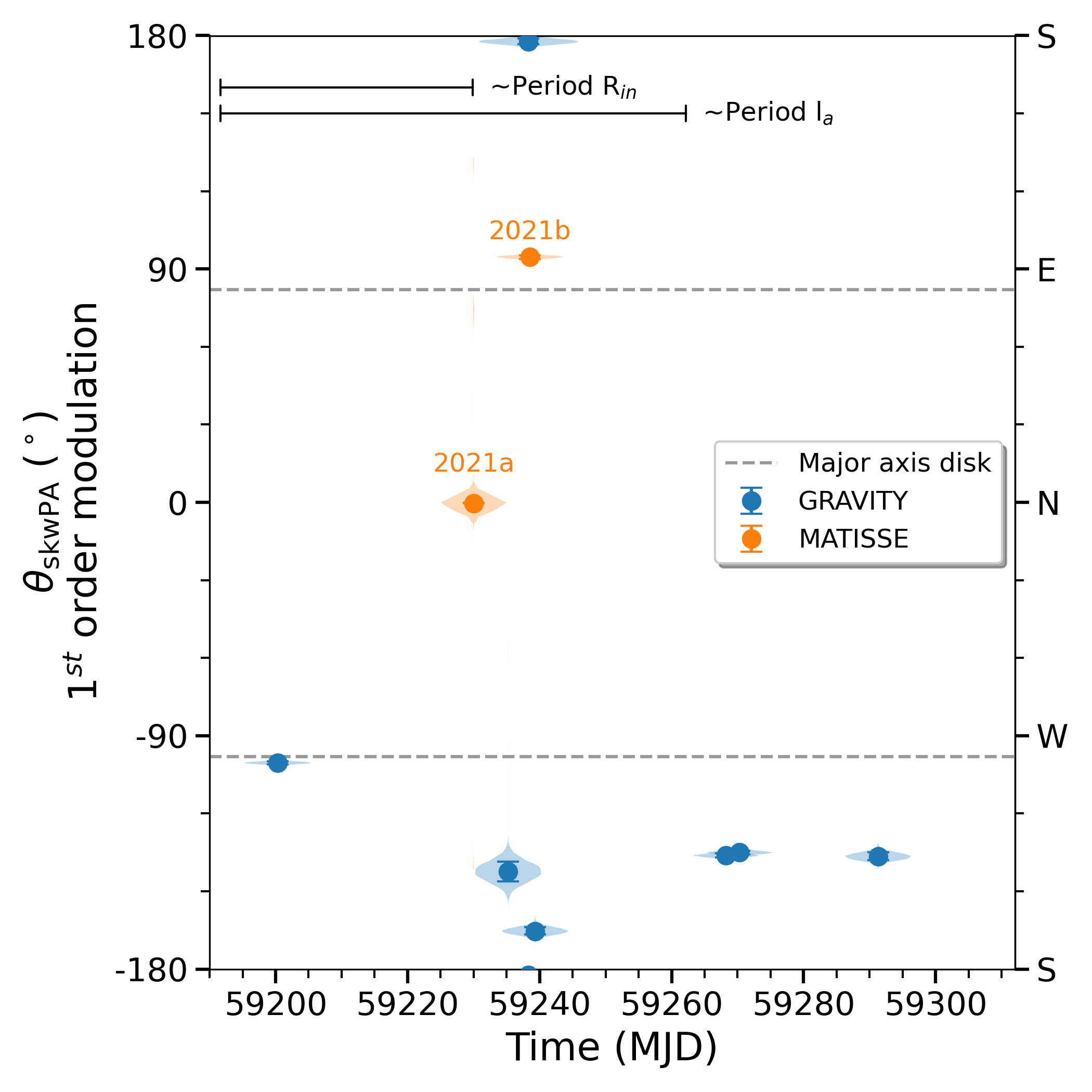}
      \caption{ MCMC results for the asymmetry angle of the first-order skewed ring model using MATISSE data from the 2021ab epoch (orange), alongside re-reduced GRAVITY data (blue) that were obtained close in time (similar to Fig. \ref{Fig:Time_Variability}). The horizontal dashed lines indicate the averaged H, K and L band estimates for the direction of the semi-major axis of the inner dust ring. In the top left corner, we indicate the approximate Keplerian period based on the stellar parameters, the best-fit temperature gradient inner radius $R_{in}$ , and the  parametric half-light radius $l_a$. 
           }
    \label{Fig:SKWPA_Time_Gravity_fit}    
\end{figure}

To first order, the inclined obscuration effect is unlikely to be the source of the observed asymmetry in general because the major axis positions are more consistent with a dynamical origin.
The L band observations were quite sparse in time, and we were therefore unable to use the L band to constrain potential motion. This analysis can be performed for the separate GRAVITY nights, however, which sample much closer in time (seven nights in about $90$ days).
Following our previous method, we first fitted a general asymmetric ring model to do this, and we then let only the asymmetry parameters $\theta_{\rm{skwPA}}$ and $A_{\rm{skw}}$ vary in time (see Fig. \ref{Fig:SKWPA_Time_Gravity_fit}). Curiously, the K band epochs showed that the preferred asymmetry angle estimates remained quite stable in the south-west over the probed timescale. This appears to contradict the estimated Keplerian motion in this region, whose period spans around $R_{\rm{in}}\approx 41\pm 0.8$ days to $l_{\rm{a}}\approx 73.9\pm 1.4$ days. 

The only two epochs that are close in time, MATISSE 2021a and 2021b, change in their best-fit asymmetry angle by $\approx 90$ degrees within only about 8 to 11 days. This timescale appears to be more consistent with Keplerian motion, but is somewhat faster. With the sparse temporal coverage, it is impossible to determine whether this shift reflects a genuine orbital evolution or observational limitations, however.
Unlike the GRAVITY epochs, the two MATISSE epochs do not probe identical spatial scales. The 2021a long baselines are more sensitive to smaller-scale variations than the medium and short baselines in 2021b. If the source morphology is intrinsically more complex than a simple first-order asymmetry, these differences in spatial sensitivity might bias the retrieved asymmetry angle. 

\subsection{Higher-order asymmetries and chromaticity as seen by MATISSE and GRAVITY} 
In Sect. \ref{subsec:reconstruction} we reconstructed an image based on the combined 2021 MATISSE snapshots, which when combined, have a decent $uv$ sampling. The observations might be explained by disk-like emission with two higher-intensity components in the two major axis directions (east and west). This second-order-like modulation agrees better with potential asymmetries caused by disk instabilities, such as a spiral, than with inclined obscuration effects. 
Because our data are limited and the four snapshots span 8 to 11 days, the features should be interpreted with some caution. Some time-smearing effects are unavoidable with VLTI imaging because the telescopes must be moved physically to fill the uv-space. Our reconstruction is probably mostly affected by the $8.5$-day difference between the large and medium arrays, which probe the smaller-scale variations. At Keplerian rotation, this gap might correspond to a potential shift of $\approx41$ degrees at the half-light radius or to a shift of up to $\approx75$ degrees at the inner rim. If the dust rotation is slower than Keplerian, however, as seen in other protoplanetary disks, the potential time-smearing might also be smaller by a factor of $\sim2$  (\citeads{2023ApJ...947...68I},\citeads{2024A&A...684A.200G},  \citeads{2025AJ....169..318S}). 

Serendipitously, two of our MATISSE/GRAVITY observation nights overlap. This allowed us to directly compare the disk asymmetry without having to account for time variability. The best-fit position angles for the first-order asymmetry between the instruments differ by about 90 to 100$^\circ$, indicating a chromatic or (probed) spatial scale effect between the K and L band. The corresponding observations are presented in closer detail side by side with simulated closure phases from the skewed ring models in Appendix \ref{App:MATISSE/GRAVITY_comparison}. 

We verified that this mismatch between the preferred direction for GRAVITY and MATISSE of the asymmetry corresponded to a real difference in the emission at these wavelengths and was not introduced by the data processing. The closure phase measurements were verified using multiple versions of the reduction pipelines: the MATISSE DRS pipeline 1.7.5 and 2.0.2, and a re-reduction for published GRAVITY data with the newer GRAVITY pipeline 1.7.0. 
Our calibrator star observations showed no anomalies either. The differential and closure phases both remained close to zero.
We also verified the same calibrator star observations by reducing and modelling a science target on the same night with known asymmetry using the same model as described in Sect. \ref{subsec:matfirst}. The asymmetry direction was consistent with the predicted asymmetry derived from models based on previous GRAVITY data (van Haastere et al., in prep). 
Taken together, this strengthens our assumption that the apparent chromatic mismatch in closure phases for MATISSE versus GRAVITY in HD~100453 is a real effect that indicates a more complex asymmetry and is not caused by an error in reduction or calibration. 

The complexity of the HD~100453 inner disk might require an approach similar to \citeads{2025AJ....169..318S} to be untangled. The authors observed the inner disk of HD 163296 at high time fidelity with the VLTI/PIONIER and the Michigan Infrared Beam Combiner (MIRC-X) at the CHARA Array and parametrically modelled the asymmetry up to eighth order. We experimented with higher-order asymmetries in our modelling to explain the K and L band data. While the MATISSE data improved in $\chi^2_r$ when higher forms of modulation were included, as also apparent from the image reconstruction, the improvement for the GRAVITY data is less clear. While the K band does not exclude higher forms of modulation, the data do not explicitly call for it. This smaller decrease in $\chi_r^2$ has to do with a smaller overall amplitude of the modulation that is seen in the K band.

The connection between the brightness of an asymmetry and closure phase (CP) is not quite intuitive \citepads{2007NewAR..51..604M}. A large phase signal might be thought to correspond to a strong asymmetry, but a relatively weak asymmetry can cause a large CP signal as well. The L band squared visibility versus baseline plot in Fig. \ref{Fig:overview_data_all} shows that the visibility function reaches zero at $\sim 30$ M$\lambda$, which means that at this baseline, the real part of the visibility $\mathfrak{Re}(V)$ becomes negative. For a centrally symmetric brightness distribution, when $\mathfrak{Re}(V)>0$, the visibility phase is 0$^\circ$, while for $\mathfrak{Re}(V)<0$, the phase is 180$^\circ$. For a visibility function without zero crossings (e.g. a Gaussian, Lorentzian, or point source), $\mathfrak{Re}(V)$ is always positive, and the closure phase is always 0$^\circ$. For any non-centrally symmetric brightness distribution, the phase can take any value between -180$^\circ$ and 180$^\circ$. For a ring-like image with an asymmetry, the phase therefore changes gradually between these extremes and does not transition instantaneously between 0$^\circ$ and 180$^\circ$ (as seen in symmetric cases with a zero crossing). 
Consequently, for certain asymmetries, significant deviations from zero phase may already be observed at baselines prior to the visibility zero. 

When we connect this to the measured CP signals, we also note that although the PIONIER and GRAVITY datasets span larger $B/\lambda$ (i.e. resolve smaller spatial scales than MATISSE), their measured closure phases are much smaller ($<10$ deg). This can be explained by the significantly higher stellar flux ratio in the NIR that causes $\mathfrak{Re}(V)$ to remain positive at all baselines. Despite the lower angular resolution of MATISSE, the mid-infrared instrument might offer an advantage over NIR instruments in this regard. At the MATISSE wavelengths, the lower stellar flux-to-disk ratio increases the likelihood of encountering a visibility null, which leads to a significant phase jump and makes asymmetries more detectable. 

Possible explanations for a complex asymmetry in HD~100453 are plentiful and extend from an accretion spiral (tidal wake) that bridges the gap between the inner and outer disk (\citeads{2017ApJ...848L..11K}, \citeads{2018ApJ...859..118B}) and asymmetries driven by one or more undetected planetary companions inside the cavity between the inner and outer disk to disruptions in the inner disk from magnetospheric accretion \citepads{2024MNRAS.528.2883Z}. More stationary-like dust features along the major axis might be caused by eccentric disk traffic-jam effects  \citepads{2013A&A...553L...3A}, but this fails to explain the chromatic or spatial scale shift. 
Future studies aiming to solve this mystery would greatly benefit from a simultaneous focused imaging campaign with both GRAVITY and MATISSE on similar baselines to reduce the dependence on the model assumptions.

\section{Summary}
\label{section:conclusions}
We investigated the misaligned inner disk around the Herbig Ae star HD~100453 using new L band VLTI/MATISSE interferometric observations. We used a combination of analytical models and image reconstruction to probe the disk structure and its physical properties. Our main focus points were to examine a previously reported asymmetry near the major axis and to perform a multi-wavelength comparison using archival VLTI PIONIER and GRAVITY data. Our main findings are listed below.
\begin{enumerate}
    \item 
    Following previous work in K band, we parametrically modelled the disk as a first-order azimuthally modulated ring. Our L band modelling found an inner disk with an inclination of $\approx 47.5^\circ$ and a position angle of $\approx 83.6^\circ$, which agrees with previous estimates at near-infrared wavelengths and further corroborates the evidence of a strongly misaligned inner-outer disk. 

    \item
    The $\chi^2_r$ was clearly improved when the asymmetry angle and modulation strength were free to vary with each epoch. The non-overlapping confidence intervals might indicate potential time variability, although we currently lack sufficient time-sampled data to properly constrain material movement and/or an induced brightness change caused by changing illumination conditions through the inner disk.

    \item  
    By modelling the L band together with archival H and K band data using a temperature gradient model, we showed a possible physical interpretation of the emission. A temperature-gradient disk with an inner radius of $\approx~0.27$~au, a vertical optical depth of $\tau_{\rm{NIR}} \approx 0.1-0.06$, and inner disk densities of about $\Sigma_0 \approx 10^{-3.2}$~g/cm$^2$  explains the visibilities well. A high-quality N-band UT observation would be invaluable to better constrain the dust mass, composition, grain sizes, and crystallinity in this system.

    \item 
    By comparing same-night L band and K band observations, we found that the best-fit first-order asymmetry angles differ by about $90$ to $100^\circ$. This chromatic/spatial-scale difference indicates that higher-order asymmetric structures (e.g. a spiral or other substructures) are likely needed to explain the closure phase signals. This interpretation was further supported by the MATISSE snapshot image reconstruction, which preferred a two-component asymmetric morphology rather than a simple first-order asymmetry. Further observations are needed to confirm these findings.
    
\end{enumerate}
The inner disk is complex and might vary rapidly, and snapshot observations only sparsely sample the available $uv$ space in time. We therefore recommend that future observing strategies focus on targeted imaging campaigns, preferably close in time on both MATISSE and GRAVITY to reduce the reliance on parametric models.

\section*{Data availability} 
The calibrated Oifits files are available in the JMMC Optical interferometry Database (OiDB). 

\begin{acknowledgements}
We thank the anonymous referee for their constructive and detailed comments, which helped improve the clarity and quality of this manuscript. We thank PIs Jean-Philippe Berger and Myriam Benisty for providing to their published VLTI/PIONIER and VLTI/GRAVITY calibrated data from \citeads{2017A&A...599A..85L} and \citeads{2022A&A...658A.183B} data, respectively. Based on observations collected at the European Organisation
for Astronomical Research in the Southern Hemisphere under ESO
programmes 190.C-0963, 0103.D-0153(F), 106.21JR.001, 106.21Q8.001, 106.21Q8.002, 106.21Q8.004, 108.225V.003, 108.225V.009, 110.2474.008.
The open-source \texttt{Oimodeler} package used in this research is developed with support from the VLTI/MATISSE consortium and the ANR project MASSIF, and we would like to thank the whole development team with an explicit mention to Anthony Meilland, Marten Scheuck and Alexis Matter for their work.
This research has made use of the Jean-Marie Mariotti Center OiDB and \texttt{OImaging} service part of the European Commission's FP7 Capacities programme (Grant Agreement Number 312430), available at \url{http://oidb.jmmc.fr}.
MATISSE is consortium composed of institutes in France (J-L Lagrange Laboratory, INSU-CNRS, Côte d’Azur Observatory, the University of Nice Sophia-Antipolis), Germany (MPIA, MPIfR, and the University of Kiel), the Netherlands (NOVA and the University of Leiden), and Austria (the University of Vienna). The Konkoly Observatory and the University of Cologne have also provided support in manufacturing the instrument.
J. Varga is funded from the Hungarian NKFIH OTKA projects no. K-132406, and K-147380. J. Varga acknowledges support from the Fizeau exchange visitors programme. The research leading to these results has received funding from the European Union’s Horizon 2020 research and innovation programme under Grant Agreement 101004719 (ORP). AM, BL, JCA, ML and PP acknowledge financial support from the Programme National de Physique Stellaire (PNPS) of CNRS-INSU in France.
This work made use of \texttt{Astropy}, a community-developed core Python package and an ecosystem of tools and resources for astronomy \citepads{2022ApJ...935..167A}. During the writing process a large language model was used to enhance readability. \\

\end{acknowledgements}

\bibliography{bibtex.bib}

\begin{appendix}
\section{Overview observations}
\label{appendix:observations}

Table \ref{tab:observations} shows the night conditions and observation information.
An overview of the reduced visibilities and closure phases split by measurement is shown in Fig. \ref{Fig:observations1}. \\

\begin{figure*}[!b]
   \centering
   \includegraphics[trim={2cm 0cm 2cm 1cm},clip, width=17cm]{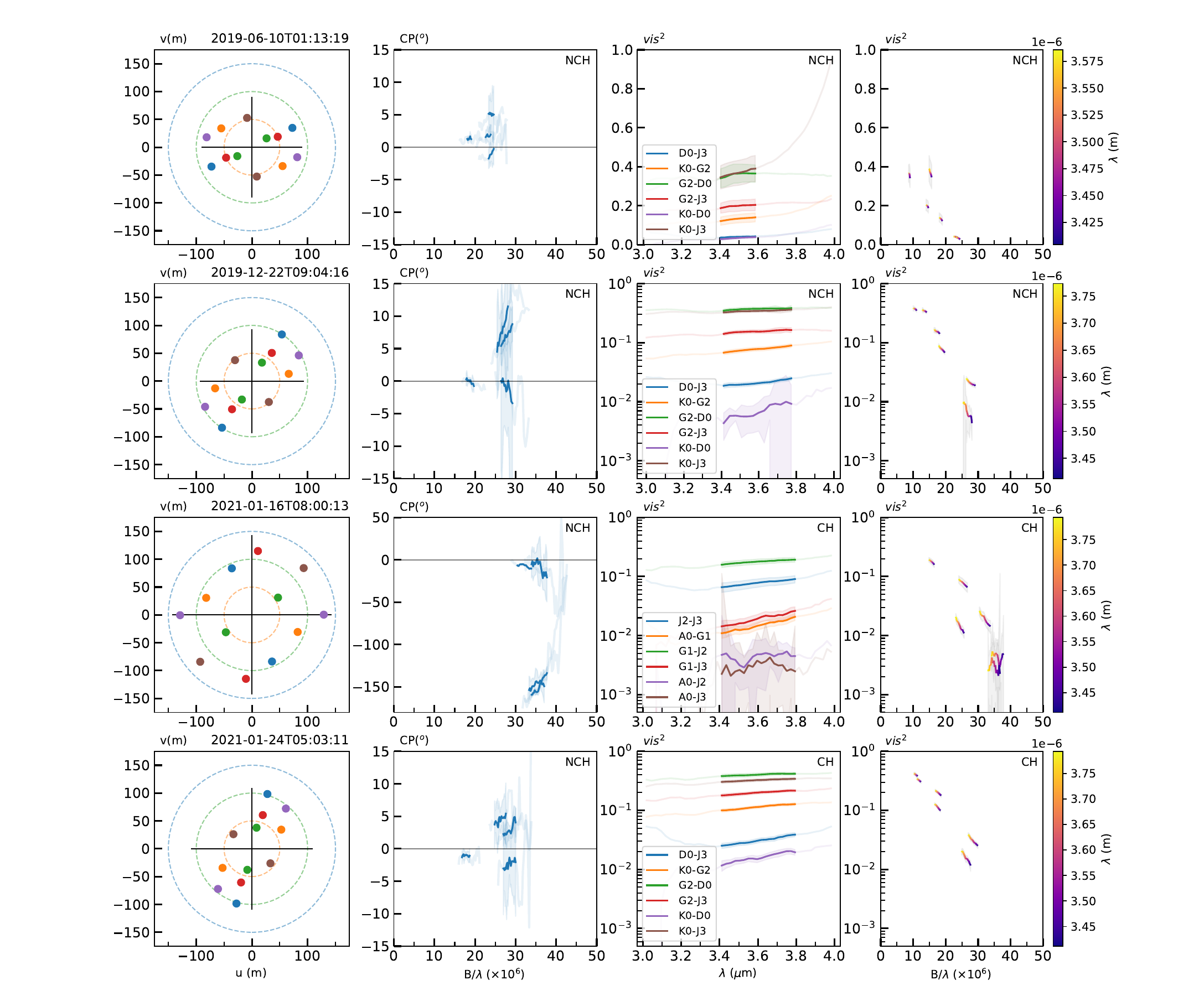}
      \caption{ Overview of all selected HD~100453 MATISSE observations, with the selected wavelength ranges highlighted. From left to right: $uv$ space, closure phase, squared visibility vs wavelength, squared visibility vs baseline length. The `CH' and `NCH' in the corner indicates whether the observable is taken from the chopped or non-chopped sequence. See Table \ref{tab:observations} for detailed night conditions. 
           }
         \label{Fig:observations1}
\end{figure*}

\begin{figure*}
   \centering
   \includegraphics[trim={2cm 0cm 2cm 1cm},clip, width=17cm]{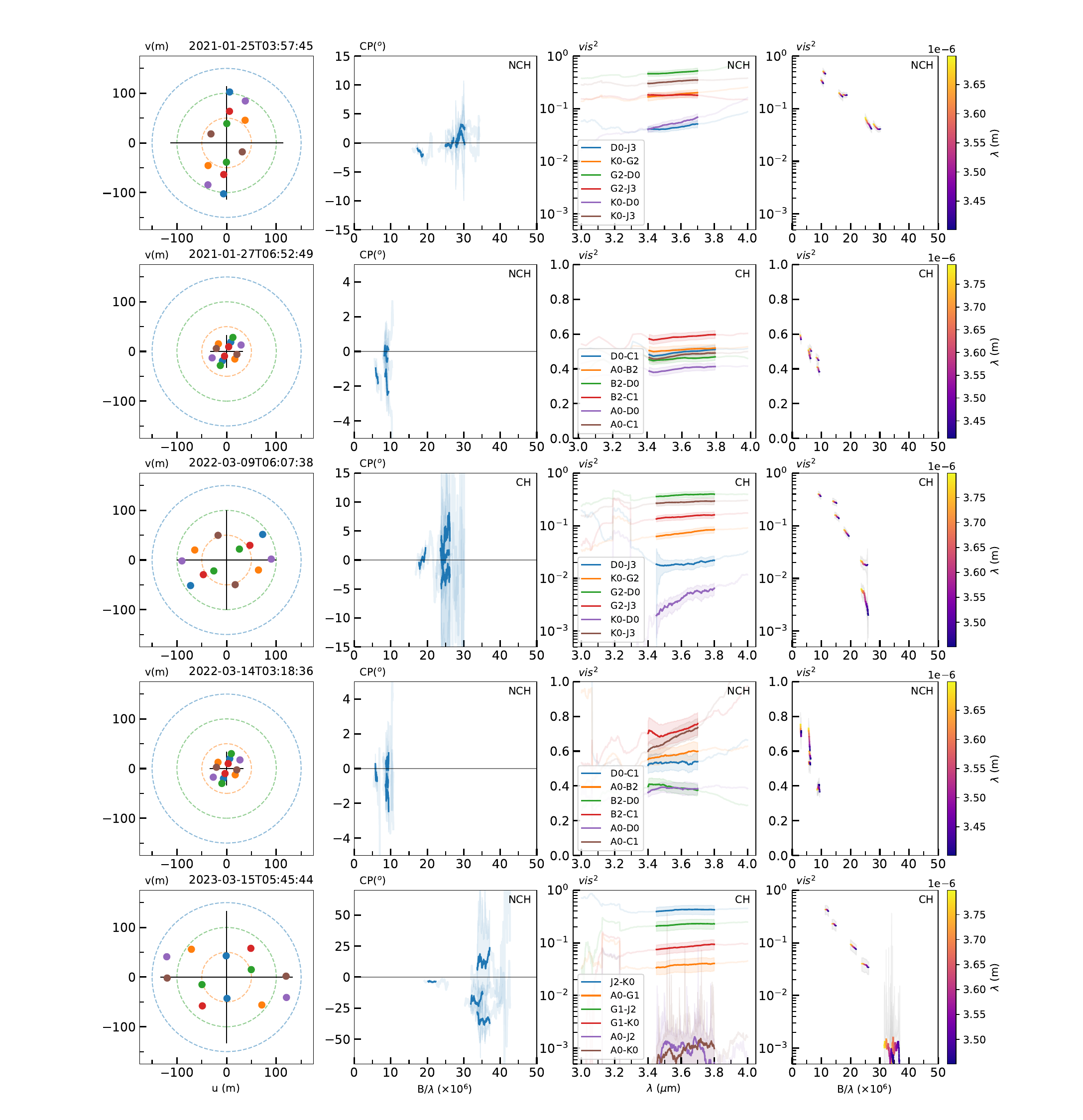}
      \addtocounter{figure}{-1}
      \caption{Continued.}
         \label{Fig:observations2}
\end{figure*}

\begin{landscape}
\begin{table}

    \caption{Overview of all HD~100453 MATISSE observations. }

        \begin{tabular}{llllllllccc|lll}
        \hline \hline
        \multicolumn{11}{c|}{HD~100453 \rule{0pt}{3ex}}                                  & \multicolumn{3}{c}{Calibrator}           \\ \hline 
                        &         &             &      &          &          &       &      &    &     &               &              &       &       \\
        Time$_{start}$  & Seeing  & $\tau_{0}$  & Wind & PWV      & Stations & Array & Res. & Ch.& G4M &  Epoch & $\Delta$Time & Star  & LDD   \\ 
             (UTC)      &  (")    & (ms)        &(m/s) & (mm)     &          &       &      &    &     &               &      (hh:mm) &       &  (mas)\\
                        &         &             &      &          &          &       &      &    &     &               &              &       &       \\ \hline
                        &         &             &      &          &          &       &      &    &     &               &              &       &       \\

                2019-05-15T02:01:55\tablefootmark{a}  & 1.45 & 2.8 & 13.8 & 2.3 & U1-U2-U3-U4 & UTs & MED &  &  & \xmark  & $+$0:34 & HD102461 & 2.97 \\ 
                2019-05-15T02:14:53\tablefootmark{a}  & 1.58 & 2.7 & 14.1 & 2.2 & U1-U2-U3-U4 & UTs & MED & \cmark &  & \xmark  & $+$0:34 & HD102461 & 2.97 \\ 
                2019-06-10T01:28:03 & 0.49 & 4.6 & 3.1 & 1.8 & K0-G2-D0-J3 & Medium & LOW &  &   & 2019a & $-$0:19 & HD102839 & 2.02 \\ 
        2019-12-22T09:11:35 & 0.67 & 11.1& 1.2 & 2.63& K0-G2-D0-J3 & Medium & LOW &  &   & 2019b & $-$0:54 & *DCen & 2.2 \\
                2021-01-16T08:09:24 & 1.18 & 6.2 & 4.6 & 2.5 & A0-G1-J2-J3 & Large & LOW &  & \cmark  & 2021a & $+$1:00 & HD102461 & 2.97 \\ 
                2021-01-16T08:20:06 & 0.96 & 5.2 & 5.1 & 2.5 & A0-G1-J2-J3 & Large & LOW & \cmark &   & 2021a & $+$1:00 & HD102461  & 2.97 \\ 
                2021-01-24T05:10:34 & 0.66 & 5.7 & 2.1 & 6.9 & K0-G2-D0-J3 & Medium & LOW &  &   & 2021b & $-$0:26 & HD92436  & 3.07 \\ 
                2021-01-24T05:17:49 & 0.55 & 7.5 & 2.0 & 6.9 & K0-G2-D0-J3 & Medium & LOW & \cmark &   & 2021b & $-$0:26 & HD92436  & 3.07 \\ 
                2021-01-25T04:06:05 & 0.66 & 5.8 & 7.6 & 4.3 & K0-G2-D0-J3 & Medium & MED &  & \cmark  & 2021b & $-$0:22 & HD92436 & 3.07 \\ 
                2021-01-27T06:51:35 & 0.77 & 4.5 & 9.7 & 5.1 & A0-B2-D0-C1 & Small & MED &  & \cmark  & 2021b & $+$0:41 & *eCen & 2.97 \\ 
                2021-01-27T07:02:05 & 0.81 & 4.9 & 8.7 & 5.1 & A0-B2-D0-C1 & Small & LOW &  & \cmark  & 2021b & $+$0:41 & *eCen & 2.97 \\ 
                2021-01-27T07:09:12 & 0.74 & 6.0 & 8.6 & 5.1 & A0-B2-D0-C1 & Small & LOW & \cmark &   & 2021b & $+$0:41 & *eCen & 2.97 \\ 
        2022-03-09T06:15:45\tablefootmark{b} & 0.44 & 16.2 & 2.6 & 7.2 & K0-G2-D0-J3 & Medium & MED &  & \cmark & \xmark & $+$0:26 & HD102461 & 2.97 \\ 
        2022-03-09T06:23:37 & 0.53 & 12.9 & 2.8 & 7.34  & K0-G2-D0-J3 & Medium & MED & \cmark & \cmark  & 2022 & $+$0:26 & HD102461 & 2.97 \\ 

                2022-03-14T03:26:38 & 1.08 & 3.4 &  7.4 & 6.3 & A0-B2-D0-C1 & Small & MED &  & \cmark  & 2022 & $+$0:27 & HD91942 & 3.49 \\ 
                2022-03-14T03:38:44\tablefootmark{c}  & 0.95 & 4.0 & 7.6 & 6.7 & A0-B2-D0-C1 & Small & MED & \cmark & \cmark & \xmark  & $+$0:27 & HD91942 & 3.49 \\ 
                2022-12-18T07:56:01\tablefootmark{a}  & 1.6 & 1.9 & 10.5 & 1.2 & A0-G1-J2-K0 & Large\tablefootmark{$\dag$} & LOW &  &  & \xmark  & $+$0:23 & HD102461 & 2.97 \\ 
                2022-12-18T08:05:33\tablefootmark{a}  & 1.57 & 2.3 & 11.0 & 1.2 & A0-G1-J2-K0 & Large\tablefootmark{$\dag$} & LOW & \cmark &  & \xmark  & $+$0:23 & HD102461 & 2.97 \\ 
                2023-03-15T05:54:24 & 0.66 & 5.3 & 4.9 & 11.8 & A0-G1-J2-K0 & Large\tablefootmark{$\dag$} & MED &  & \cmark  & 2023 & $+$0:27 & HD102964 & 2.45 \\ 
                2023-03-15T06:05:01 & 0.44 & 9.4 & 4.7 & 11.9 & A0-G1-J2-K0 & Large\tablefootmark{$\dag$} & MED & \cmark & \cmark &  2023 & $+$0:27 & HD102964 & 2.45\\
                        &         &             &      &          &          &       &      &    &      &        &              &          &  \\ \hline

        \end{tabular}
        
        \tablefoot{
        The night information describes the weather conditions at the start of observations, where $\tau_{0}$ is the coherence time, 'Res.' the spectral resolution, 'Ch.' the chopping indicator and 'G4M' the GRAVITY fringe tracker indicator. The spectral resolution varied between observations from low to medium resolution. Six datasets were not included in the analysis after careful examination. \\
        \tablefoottext{a}{Observations with seeing $> 1.2"$ are generally considered unreliable and hence were excluded from the analysis.}\\
        \tablefoottext{b}{Rare instance where non-chopped closure phases were noisier than chopped.}\\
        \tablefoottext{c}{Excluded since conditions were not sufficient to keep fringe tracking stable during the chopping sequences.}\\
        \tablefoottext{$\dag$}{In earlier periods this array was called astrometric.}
        }

        \label{tab:observations}
\end{table}
\end{landscape}

\FloatBarrier

\section{Highlight modelled asymmetry}
\label{appendix:highlight_asymm}
The asymmetry in a skewed ring model can be difficult to clearly discern in the image plane when visualized with a monotonically increasing colour map. To enhance its apparentness, we used the following method to isolate the asymmetric component:
\begin{align}
    \label{eq:highlight_asymm}
    I_{\nu,\rm{diff}} &=  I_\nu(\rm{fit}) -  I_\nu(\rm{fit}\big\rvert_{\rm{A_{\rm{skw}}=0}})   \\
    I_{\nu, \rm{norm}} &=  I_\nu(\rm{fit})\big\rvert_{\rm{at \;argmin( I_{\nu,\rm{diff}} ) }} \nonumber \\
     \Delta I_{\nu, \rm{asymm}} &= (1-I_{\nu, \rm{norm}})\cdot\left( I_\nu(\rm{fit})  - \left(\frac{I_{\nu,\rm{diff}} }{\min(I_{\nu,\rm{diff}} )}\cdot I_{\nu, \rm{norm}}\right) \right), \nonumber
\end{align}

where $I_\nu(\mathrm{fit})$ denotes the image plane of a fitted model, and $I_{\nu, \mathrm{norm}}$ is the normalized intensity at the point on the ring located $180^\circ$ opposite the peak intensity. This position defines the new zero point in the resulting asymmetry image. 

\section{Spectral energy distribution}
\label{appendix:sed}

\begin{figure}
   \centering
   \includegraphics[width=0.5\textwidth]{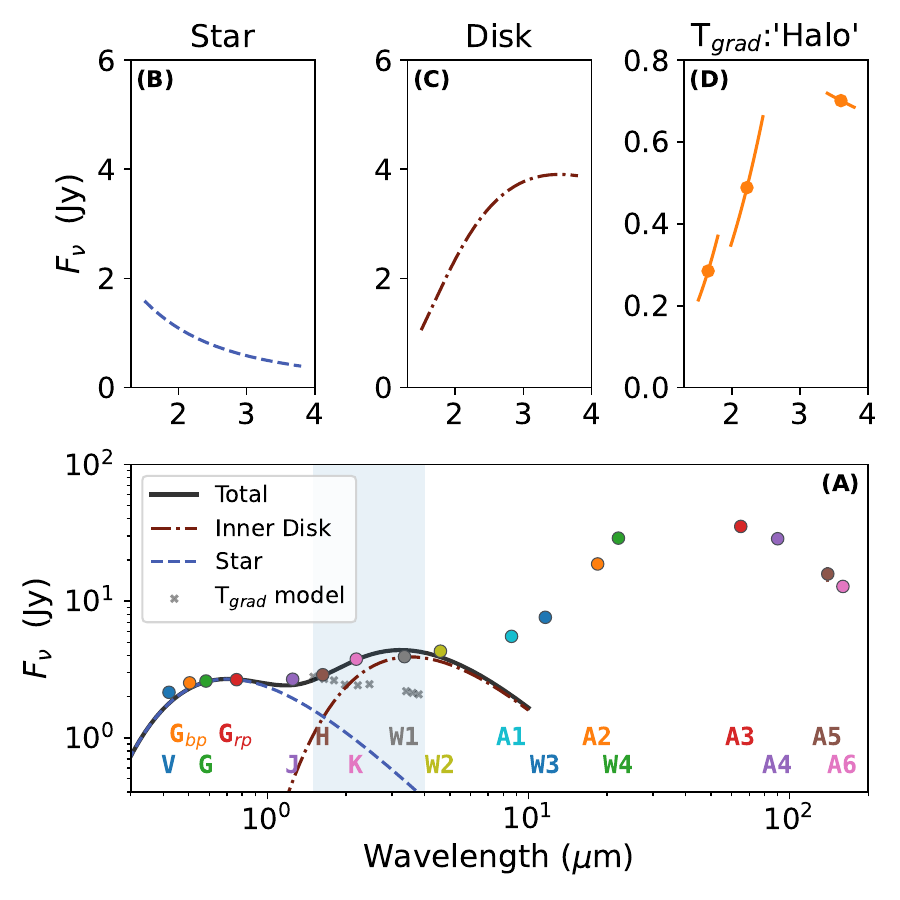}
      \caption{Panel     A: Spectral energy distribution of HD~100453 compiled from various facilities. Panels B and C: Fitted blackbody curves corresponding to the central star (B) and inner disk (C), respectively, using the stellar parameters from \citeads{2021A&A...650A.182G}. The flux from the best-fit temperature gradient model is also included (see Table \ref{tab:model_temperature_gradient}). Panel D: Contribution from the extended `halo' component. The blue highlighted area corresponds to the wavelength range probed in this research.}
         \label{FigStellarFlux2}
\end{figure}

To compare our results with photometric measurements we collected the V, G$_{BP}$, G, G$_{RP}$, J, H, K, WISE (W1-W4) and AKARI (A1-A6) band photometric data from various surveys; Hipparcos \citepads{1997ESASP1200.....E}, GaiaDR3 \citepads{2023A&A...674A...1G}, 2MASS \citepads{2006AJ....131.1163S}, WISE (\citeads{2012yCat.2311....0C}, \citeads{2021ApJS..253....8M}), AKARI (\citeads{2010A&A...514A...1I}, \citeads{2010yCat.2298....0Y}). Since the extinction $A_{\rm V}$ is approximately zero the data did not need to be dereddened \citepads{2018A&A...620A.128V}. Figure \ref{FigStellarFlux2} displays the spectral energy distribution of HD~100453 together with a star and a single-temperature  blackbody disk fit to the J, H, K, W1, and W2 photometric bands. 
We assumed the literature values for HD~100453 (see Table \ref{tab:system_parameters}) and fitted the SED ($\lambda < 5~\mu \rm{m}$) with

\begin{align}
   F_\nu & \approx B_{\nu\rm{, star}} \Omega_{\rm{star}} + B_{\nu\rm{, disk}} \Omega_{\rm{disk}}  \\
         & \approx \pi B_{\nu\rm{, star}} \left( \frac{R_{\rm{star}}}{d} \right)^2 + B_{\nu\rm{, disk}} \Omega_{\rm{disk}}, \nonumber
\end{align}

 \noindent which yielded a simple disk temperature approximation of $1445 \pm 85$ K with a scale factor of $\Omega_{\rm{disk}} \approx (6.83 \pm 1.41 )\times10^{-17}$ sr. For these values at $\sim3.5~\rm{\mu m}$ the star/disk flux ratio equals roughly $\approx0.11$. 
\\

\FloatBarrier

\section{Reliability of uncertainties in infrared interferometry}
\label{appendix:error}
\begin{figure*}
   \centering
   \includegraphics[trim={0cm 0 0cm 0},clip, width=16cm]{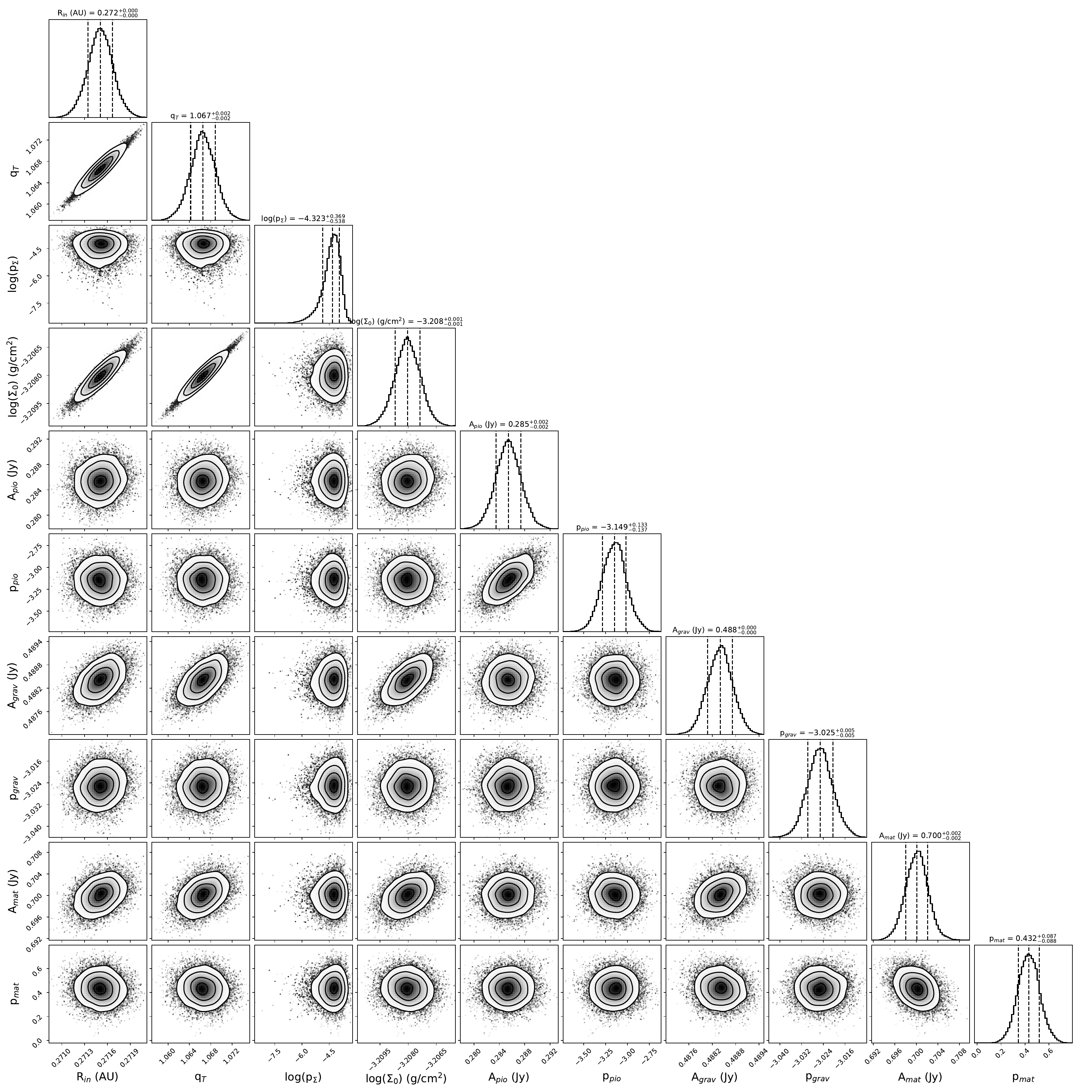}
      \caption{Temperature gradient model best-fit MCMC chains, for $R_{\rm{out}} = 10$ AU, after removing burn-in to the combined PIONIER, GRAVITY \& MATISSE datasets. N$_{walkers} = 25$, N$_{steps} = 2000$ after burn-in; see Table \ref{tab:model_temperature_gradient}.
           }
         \label{Fig:mcmc_corner_tgrad}
\end{figure*}

Figure \ref{Fig:mcmc_corner_tgrad} shows the corner-plot of one of the MCMC fits which was used for Table \ref{tab:model_temperature_gradient}. Typically IR interferometry reports the error estimate on fitted parameters as the $16\% - 84\%$ intervals from settled MCMC chains. An important underlying assumption is that this assumes the measurement errors are independent and Gaussian distributed. However, this method is flawed as interferometric observables, especially when working with multiple spectral channels, are often highly correlated and subject to systematic errors. 
For example the visibilities for different spectral channels on a baseline are affected by similar atmospheric and instrumental effects in their signal,
which can shift visibilities sets of affected baselines as a whole up and down. Similarly the visibilities and closure phases sharing a telescope or baseline are also affected by correlations due to having a common signal \citepads{2020A&A...644A.110K}.

The calibrator stars used in this research have a mean limb-darkened diameter (LDD) uncertainty of 0.6\% \citepads{2019MNRAS.490.3158C}, hence our disk fit size uncertainties on $\epsilon_{l_a} / l_a \approx0.23\%$ and $\epsilon_{R_{\rm{in}}} /R_{\rm{in}}\approx0.37\%$ respectively are rather optimistic. When interpreting fit results it is important to be cautious about the likely underestimated errors, and pay attention to trends in the visibility derivative or \mbox{residuals}, as those are less affected by this shift.
Future fitting procedures could be improved by taking this correlation into account (Priolet et al., submitted), or using other methods such as bootstrapping to better estimate the uncertainties in model parameters \citepads{2019MNRAS.484.2656L}.

\section{MATISSE/GRAVITY comparison of closure phases on the same nights}
\label{App:MATISSE/GRAVITY_comparison}

\begin{figure*}
   \centering
    \includegraphics[trim={0cm 0cm 0cm 3cm},clip,width=16cm]{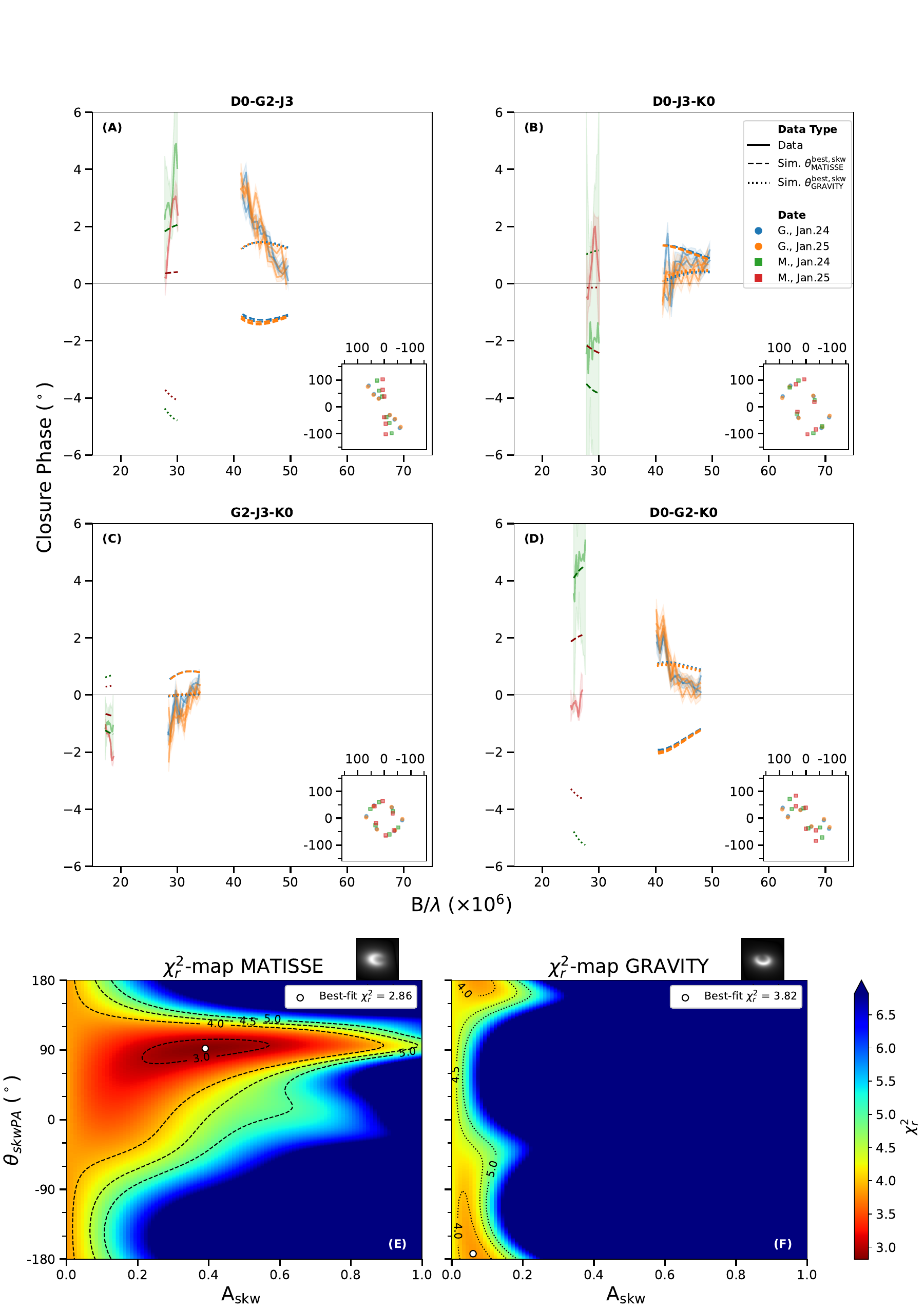}
      \caption{ Direct comparison of the GRAVITY and MATISSE closure phase observations taken on two overlapping nights, 24 and 25 January 2021 ($\approx59238.5 $ MJD). Panels A--D: Measurements for the four sets of baseline triangles, with an inset plot in the corner showing the corresponding ($u,v$) coordinates in meters. The solid line shows our binned data, while the dashed and dotted line show the simulated data of the best-fitted skewed ring for these dates with corresponding $\theta_{\rm{skwPA}}$ for MATISSE ($\approx 92^\circ$) and GRAVITY ($\approx -173^\circ$) respectively; see Fig. \ref{Fig:SKWPA_Time_Gravity_fit}. Panels E and F: The $\chi_{\rm{r}}^2$-maps for the fitted asymmetry parameters, together with a representation of the best-fit direction for the respective bands. 
           }
         \label{Fig:mat2425_grav_2425}
\end{figure*}

Figure \ref{Fig:mat2425_grav_2425} shows a comparison of how the simulated closure phases for the skewed ring look in practice. From both visual inspection and the $\chi^2_{\rm{r}}$-maps it is clear that the $\theta_{\rm{skwPA}} \approx 92^\circ$ represents the MATISSE data much better than the best fit value $\theta_{\rm{skwPA}} \approx -173^\circ$ from the GRAVITY fit, and vice versa for the GRAVITY data.

\end{appendix}

\end{document}